# Broadband Light Harvesting from Scalable Two-Dimensional Semiconductor Heterostructures

Da Lin[1], Jason Lynch[2], Sudong Wang[2], Zekun Hu[2], Rajeev Kumar Rai[1], Huairuo Zhang[3,4], Chen Chen[5], Shalini Kumari[5], Eric Stach[1], Albert V. Davydov[3], Joan M. Redwing[5, 6], Deep Jariwala[1,2]

[1] Materials Science and Engineering, University of Pennsylvania, Philadelphia, PA, USA

[2] Electrical and Systems Engineering, University of Pennsylvania, Philadelphia, PA, USA

[3] Materials Science and Engineering Division, National Institute of Standards and Technology, Gaithersburg, MD, USA

[4] Theiss Research, Inc., La Jolla, CA 92037, USA

[5] 2D Crystal Consortium Materials Innovation Platform, Materials Research Institute, Penn State University, University Park, PA, USA

[6] Materials Science and Engineering, Penn State University, University Park, PA, USA

## Abstract

Broadband absorption in the visible spectrum is essential in optoelectronic applications that involve power conversion such as photovoltaics and photocatalysis. Most ultrathin broadband absorbers use parasitic plasmonic structures that maximize absorption using surface plasmons and/or Fabry-Perot cavities, which limits the weight efficiency of the device. Here, we show the theoretical and experimental realization of an unpatterned/planar semiconductor thin-film absorber based on monolayer transition metal dichalcogenides (TMDCs). We experimentally demonstrate an average total absorption in the visible range (450 nm – 700 nm) of > 70% using < 4 nm of semiconductor absorbing materials scalable over large areas with vapor phase growth techniques. Our analysis suggests that a power conversion efficiency (PCE) of 15.54% and a specific power > 300 W g$^{-1}$ may be achieved in a photovoltaic cell based on this metamaterial absorber.

## Introduction

Absorption occurs when light-matter interactions lead to an energy transfer from an incident photon to the irradiated material. In the context of optoelectronic devices, absorption is typically required to be either narrowband (single wavelength) or broadband (spectral) depending on the application. In the past decade, the field of thin-film, near-unity broadband absorption has garnered interest given its numerous benefits compared to bulk absorbers, such as decreased material usage and cost, shorter carrier collection distance that decreases recombination losses[1], and decreased weight – these characteristics open doors for efficient ultrathin optoelectronic applications, such as photovoltaics[2], photoelectrochemical processes[3,4], and photodetection[5].

In the submicron range, the dominant approach to achieving broadband absorption has been engineering metamaterials using plasmonic nanocavities to generate localized surface plasmon-polaritons to strongly confine light in the absorbing material[6]. Metal-insulator-metal (MIM) structures were extensively studied with various plasmonic structures as the top layer[7–10]. Unfortunately, the energy absorbed by metals is lost through the thermalization of carriers, which is not useful and often detrimental for optoelectronic devices. To combat this issue, thin-film semiconductors have been coupled to plasmonic nanostructures to enhance light-matter interaction[11–13]. However, employing metal plasmonic structures as the dominant mechanisms for strong light-matter coupling results in Joule heating losses in the metal – this parasitic absorption does not contribute to the optoelectronic function, decreasing the weight efficiency of the device. One strategy to avoid parasitic absorption in the metal nanostructures has been to pattern the semiconductor itself to form cavity modes that couple to the incident light. Although this has been done to a 1 µm Si slab to achieve > 80% absorption in the visible range[14], it comes at the expense of scalability because of the complex patterning needed. Another strategy to increase the strength of light-matter interactions in a planar geometry is to alter the thickness of a layer to create a thin film interference effect[15,16]. Absorption in planar semiconducting thin films on top of metal or metal/dielectric substrates has been demonstrated for $MoSe_2$[17], Ge[13,18], $MoS_2$[19,20], and graphene[21]. While these studies concentrate most incident light into the semiconducting active layers, they do not push the boundary of ultrathin materials – the atomically thin limit – that would achieve the greatest energy efficiency by weight.

Previous work from our team has demonstrated near-unity narrowband absorption using a scalable multilayer superlattice structure based on atomically thin 2D transition metal dichalcogenides (TMDCs) measuring < 1 nm in thickness per layer[22]. In this letter, we numerically and experimentally demonstrate a broadband visible absorber based on < 4 nm of large area, scalable, planar (unpatterned) monolayer TMDCs. We achieve electronic isolation of each semiconducting layer within the superlattice stack to obtain multiple resonance peaks, contributing to a broadband absorption spectrum with an average total (sum of the absorptance of each layer) absorption in the visible range (450 nm – 700 nm) greater than 70%. This is the thinnest, scalable broadband absorber reported to-date without the use of any bottom-up self assembly, top-down nanopatterning or integration of complex plasmonic nanostructures. We also simulate the potential photovoltaic performance of the superlattice and find that it has a large potential specific power (> 300 W $g^{-1}$). Our structure pushes the light-matter coupling efficiency limit of ultrathin semiconducting materials, applicable for a plethora of optoelectronic devices.

## Results and Discussion

**Design and Fabrication of Ultrathin TMDC Absorber**

**Figure 1a** illustrates our ultrathin broadband absorber design employing alternate layers of semiconducting (TMDC) monolayers and dielectrics ($Al_2O_3$), along with a back reflector (Ag). This geometry constitutes a multi-quantum well (MQW)[23], or superlattice (SL), and it has been reported in our work that employed single TMDC to enhance the narrowband absorption of the SL[22]. Monolayer TMDCs are chosen for the semiconductor layers because of their strong, quantum-confined excitons that are highly absorptive, their bandgaps that range from 1.1 to 2.0 eV, and their direct bandgap nature in the atomically thin limit which is highly desirable in optoelectronic applications[24–30]. Although the four bottom layers follow the trend of decreasing bandgap from top to bottom which is typical in multi-tandem solar cells[31], the smallest bandgap (2H-$MoTe_2$) is placed on the top. This is done to reduce the amount of degradation of $MoTe_2$ during fabrication since it is the most sensitive TMDC used to the ambient environment[32,33]. Simulations show negligible differences in absorption when the $MoTe_2$ is on the top or bottom of the SL since the overall thickness of the structure is much smaller than the wavelength of photons in the visible part of spectrum (Supplementary Information **Figure S1**). The $Al_2O_3$ insulating layers are 3 nm thick because this is thick enough to isolate the exciton wavefunction to a single monolayer, thus inhibiting interlayer coupling effects[34,35]. The bottom dielectric underneath the SL strongly affects the light-matter interactions over a narrowband range because of its effect on interference within the SL. Because of this, the bottom dielectric function thickness is the only degree of freedom (DoF) that is used to optimize the broadband absorption of the system. The optimization process is done using the transfer matrix method (TMM) along with refractive index values measured using spectroscopic ellipsometry (Supplementary Information **Figure S2**) to calculate the absorption of the superlattice. The refractive index of monolayer 2H-$MoTe_2$ was approximated as its bulk value (10 nm) because of a lack of the monolayer value in literature[36]. The back mirror is 200 nm of E-beam evaporated Ag, which is far thicker than the penetration of visible light (14 nm @ $\lambda$=500 nm). This ensures that no transmission occurs through the device, and absorption can be calculated directly from reflection (A = 1 - R). Silver is chosen specifically to minimize parasitic absorption[37].

One key advantage of using TMDC monolayers is the advent of techniques to isolate and deposit pristine monolayer TMDCs on any arbitrary surface on the wafer-scale[38–40], as shown in **Figure 1b**. MOCVD is used to grow large area monolayer $MoSe_2$ ($N$ = 1), $WSe_2$ ($N$ = 2), $WS_2$ ($N$ = 3) and $MoS_2$ ($N$ = 4) on sapphire for the large area superlattice. However, for the partially exfoliated superlattice, $MoSe_2$ and $WSe_2$ are exfoliated. The films are then transferred to a new substrate using the wet transfer process to fabricate the superlattice (See Methods). Monolayer $MoTe_2$ ($N$ = 5) are mechanically exfoliated from a bulk crystal using a PDMS stamp technique[40]. As demonstrated in subsequent sections, both techniques of transfer yield high-quality samples with minimal defects, with the $N$ = 4 SL demonstrating exemplary environmental stability. Atomic layer deposition (ALD) is used to deposit thin-film $Al_2O_3$ on top of the back reflector and between the monolayer TMDC layers. Supplementary Information **Figure S3** demonstrates a separate realization of the superlattice stack using mechanically exfoliated $MoSe_2$ ($N$ = 1), $WSe_2$ ($N$ = 2), and $MoTe_2$ ($N$ = 5) and MOCVD grown $WS_2$ ($N$ = 3) and $MoS_2$ ($N$ = 4). It is important to note that all of the TMDC monolayers have been grown on the wafer-scale using MOCVD, demonstrating the scalability of our system[41–43]. In Supplementary Information **Figure S4**, TEM images and EDS maps are captured, where the individual layers and identities of the superlattice stack can clearly be seen. Further details on the experimental process may be found in the **Supplementary Information**.

# Figure 1

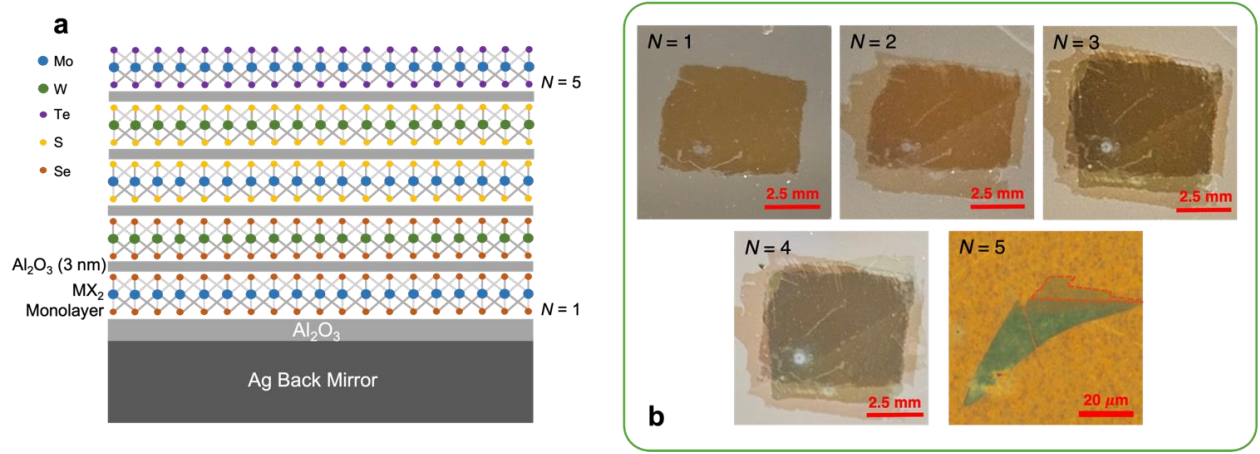

Figure 1. (a) Front-view schematic of a $N$ = 5 TMDC superlattice structure composed of TMDC monolayers, dielectric spacers ($Al_2O_3$) with a bottom dielectric thickness of 33 nm, and a silver back reflector mirror. (b) Optical microscope (OM) images of the step-by-step fabrication process of a $N$ = 5 large area sample.

## Optical Properties of TMDC Absorber

A distinctive feature of monolayer TMDCs from their multilayer counterparts is their strong photoluminescence (PL) which is a result of the indirect to direct bandgap transition in the monolayer limit[44,45]. Interlayer coupling causes multilayer TMDCs to suffer from increased non-radiative recombination that quenches the intensity of the PL, reducing their potential optoelectronic efficiency[34,46]. In **Figure 2a**, we verify the monolayer thicknesses and electronic isolation of the fabricated TMDC layers within our SL through PL measurements of a characteristic $N$ = 5, partially exfoliated SL ($WSe_2$, $MoSe_2$, and $MoTe_2$ were mechanically exfoliated while the $MoS_2$ and $WS_2$ are large-area and were transferred using a wet transfer process). The resulting peaks are generally red shifted from the A exciton resonance energy for the identified monolayer TMDC due to phonon energy loss in radiative recombination, and all of the peaks are within 50 meV of prior measurements[26–29,47]. Variations are attributed to mechanical strain introduced to the SL during the extensive fabrication process[27,48], as well as additional dielectric screening from the surrounding environment that decreases the electronic bandgap of the TMDC[49]. This is particularly relevant for the bottom layer, $MoSe_2$, that has the largest deviation from previous works while also experiencing the most mechanical mismatch, environmental contamination, and dielectric encapsulation. The blue-shift of the peak, in particular, can be attributed to a strong in-plane compressive stress[48] placed on the monolayer as a result of 8 additional layers on top of it. This blue-shift is observed throughout the fabrication process (Supplementary Information **Figure S5**). The relative intensities of the peaks ($WS_2$, $WSe_2$, $MoS_2$, $MoSe_2$) are consistent with the simulated absorption of the monolayers at 405 nm (Supplementary Information **Figure S6**) and experimentally demonstrated PLQY values[40,50–52]. The main exception to this trend is $MoSe_2$, which performs significantly worse than expected in comparison to the other TMDCs – this is attributed to increased defects states and strains introduced in the fabrication process, as outlined above, as well as the screening of photons as they travel through multiple interfaces in the SL to the collector. The $MoX_2$ (X = S, Se) peaks are also observed to have larger full-width-half-maxes (FWHMs) compared to their $WX_2$ (X = S, Se) equivalents by 27 nm and 51 nm, respectively. This behavior is expected for $MoS_2$, which inherently scatters more energy to phonons than $WS_2$[53], as well as for $MoSe_2$ where increased localized defect scattering combines with the A exciton peak[54].

We further probe the photoluminescence behaviour of our superlattice by taking low temperature (LT) measurements of the SL at 80 K, as shown in the inset of **Figure 2(a)** and deconvoluted in **Figure S7** in the Supplementary Information. The most notable shifts in PL resonance are in the $MoS_2$ peak, which experiences a red-shift of 25 meV, and $MoSe_2$, with a red-shift of 45 meV. The red-shifting is a result of the thermal expansion coefficient mismatch between layers changing the stress in the TMDC layers and counteracting the typical effects of cooling.

The WS$_2$ and WSe$_2$ experience negligible shifts. These peaks shift as temperature decreases, and this can be attributed to increased thermally induced tensile stress[27]. MoSe$_2$ experiences the largest thermally induced tensile stress because it has the highest coefficient of linear thermal expansion among the TMDCs and dielectric materials, while MoS$_2$ is the second highest[55–59]. Conveniently, the defects in MoSe$_2$ caused by the initial compressive stress at 300 K is relieved as the temperature is decreased, increasing the relative intensity of the MoSe$_2$ PL compared to WSe$_2$.

The identity and thickness of the monolayers are further confirmed using Raman spectroscopy **(Figure 2b)**, which plots the Raman spectra for the $N$ = 5 SL against experimentally acquired Raman spectra for CVD-grown or mechanically exfoliated monolayers. Thickness-dependent Raman spectra is highly accurate in pinpointing layer-by-layer increases in thickness due to additional interlayer vdW interactions that stiffen the $A_{1g}$ mode from the increased vibration restoring force, as well as stacking effects that relax the $E_{2g}$ mode from shifting lattice dimensions[60,61]. As seen, the top two layers are easily discernible in the $N$ = 5 spectra – the $A_{1g}$ mode for monolayer MoTe$_2$ and WS$_2$ closely match that of previously acquired values[30,61]. The bottom layers are more difficult to identify due to background noise attributed to interfacial screening; however, layer-by-layer measurements of an $N$ = 3 SL are clearly identified $A_{1g}$ and $E_{2g}$ peaks corresponding to monolayer values (Supplementary Information **Figure S8**)[29,62].

**Figure S9** demonstrates equivalent data in the representative $N$ = 5, large area SL depicted in **Figure 1b**, where electronic isolation is clearly demonstrated in the photoluminescence and Raman measurements.

# Figure 2

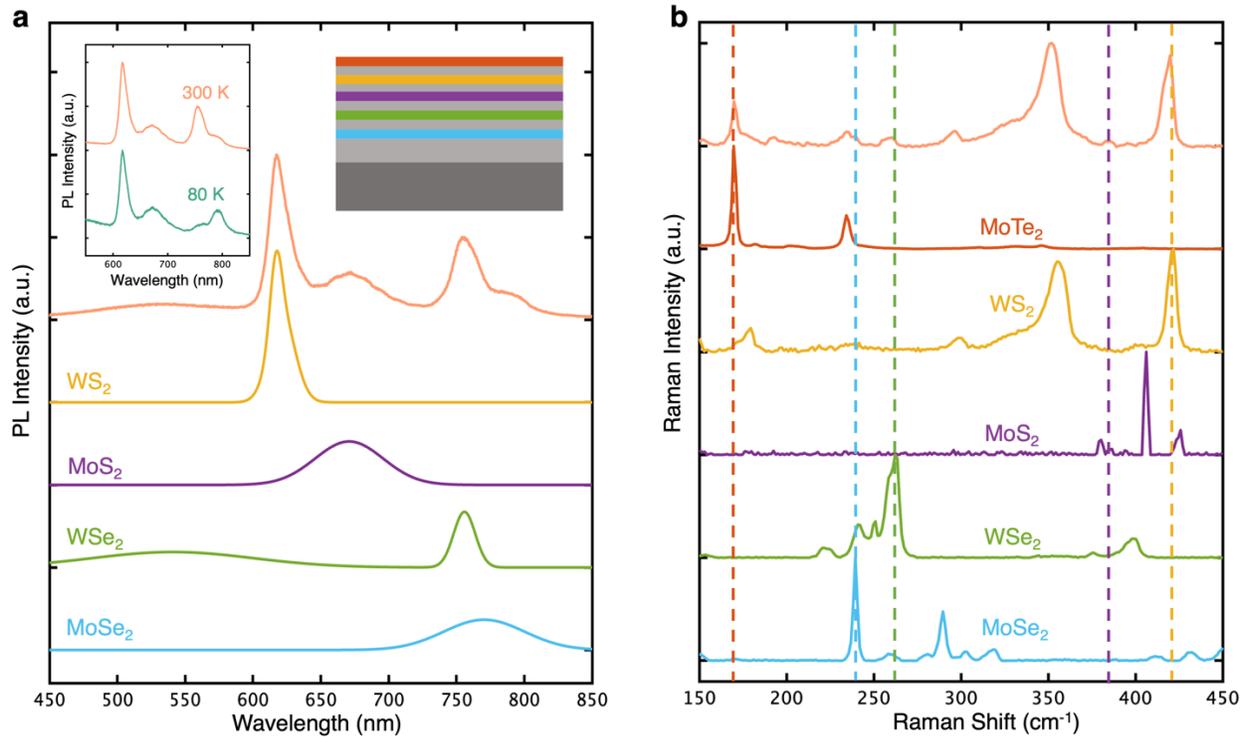

Figure 2. (a) Normalized photoluminescence (PL) spectra for the superlattice (orange) and deconvoluted Gaussian fits for individual layer PL spectra for the partially exfoliated superlattice. (b) Normalized Raman spectra for superlattice matched with experimental Raman spectra for CVD-grown/dry-transferred TMDC monolayer. Inset of (a): (Left) Temperature-dependent PL spectra for superlattice at 300 K and 80 K. (Right) Simplified front-view schematic of superlattice structure.

The absorption spectra are obtained experimentally for a characteristic $N = 4$ and $N = 5$ SL for both large area and partially-exfoliated samples (**Figure 3a and 3b**). Absorption is calculated as 1 – Reflection, given that the Ag back reflector blocks all transmission. The multilayer reflections within the SL lead to a broadband, multi-resonant peak absorption spectrum within the visible range (450 nm – 700 nm). An average total absorption of 58.5% and 73.8% is achieved for the partially-exfoliated $N = 4$ (active thickness = 2.8 nm) and $N = 5$ SL (active thickness = 3.5 nm), respectively, while an average total absorption of 60.6% and 70.5% is achieved for the large area $N = 4$ and $N = 5$ SL. Further, in the 450 nm – 520 nm range, the partially-exfoliated $N = 5$ SL demonstrates > 80% average total absorption while the large area N = 5 SL demonstrates > 75% average total absorption. The broadband absorption is attributed to a combination of the strong excitonic resonance peaks of the individual semiconducting layers and an enhancement of light-matter interactions from the optimized thickness of the bottom dielectric. The first resonant peak at 516 nm is attributed to the B exciton peak of $WS_2$[63]. A and B exciton peak splitting is well-documented and is attributed to the splitting of the valence band due to spin-orbit coupling[64]. The second resonance peak at 614 nm is predominately the A exciton peak of $WS_2$ and the B exciton peak of $MoS_2$, while the third resonance peak at 653 nm corresponds to the A exciton of $MoS_2$. **Figure S10** in the Supplementary Information additionally shows experimental resonance peaks at 745 nm from the $WSe_2$ A exciton and 788 nm and 704 nm from the $MoSe_2$ A and B exciton in the partially exfoliated sample. Solar absorption under 1.5 AM solar irradiation is calculated in **Figure S11**.

The absorption of our SL is simulated using the TMM[65] to verify that our experimental results are consistent with theoretical models. The simulated absorption spectra in **Figures 3a and 3b** were found to agree well with partially-exfoliated sample, while variations in the large area sample are attributed to introduced impurities and defects in the growth and wet transfer process. We verify that the broadband visible absorption of the SL has been optimized using a constrained minimization algorithm where the bottom dielectric thickness $t$ is varied, resulting in $t = 33$ nm for the $N = 4$ superlattice and $t = 31$ nm for the $N = 5$ superlattice. For the sake of experimental fabrication, 33 nm alumina was deposited for the sample device presented in this text.

The TMM model for our SL allows for the layer-by-layer analysis of our structure **(Figure 3c and 3d)**. The average total broadband absorption, in descending order, of the $N = 4$ (68.77%) model is $MoSe_2$ (25.28%), $WSe_2$ (17.79%), $MoS_2$ (10.56%), $WS_2$ (9.38%), and Ag (5.76%) while the $N = 5$ (79.19%) model is $MoTe_2$ (21.37%), $MoSe_2$ (21.03%), $WSe_2$ (14.85%), $MoS_2$ (8.93%), $WS_2$ (7.94%), and Ag (5.06%). Layer absorption is particularly enhanced in the bottom layer of the SL, where absorption values surpass the values expected from freestanding, electronically isolated monolayers[66,67]. A close fit exists between the experimental result and the layer-by-layer predicted excitonic resonances, which further confirms the consistency of the SL.

# Figure 3

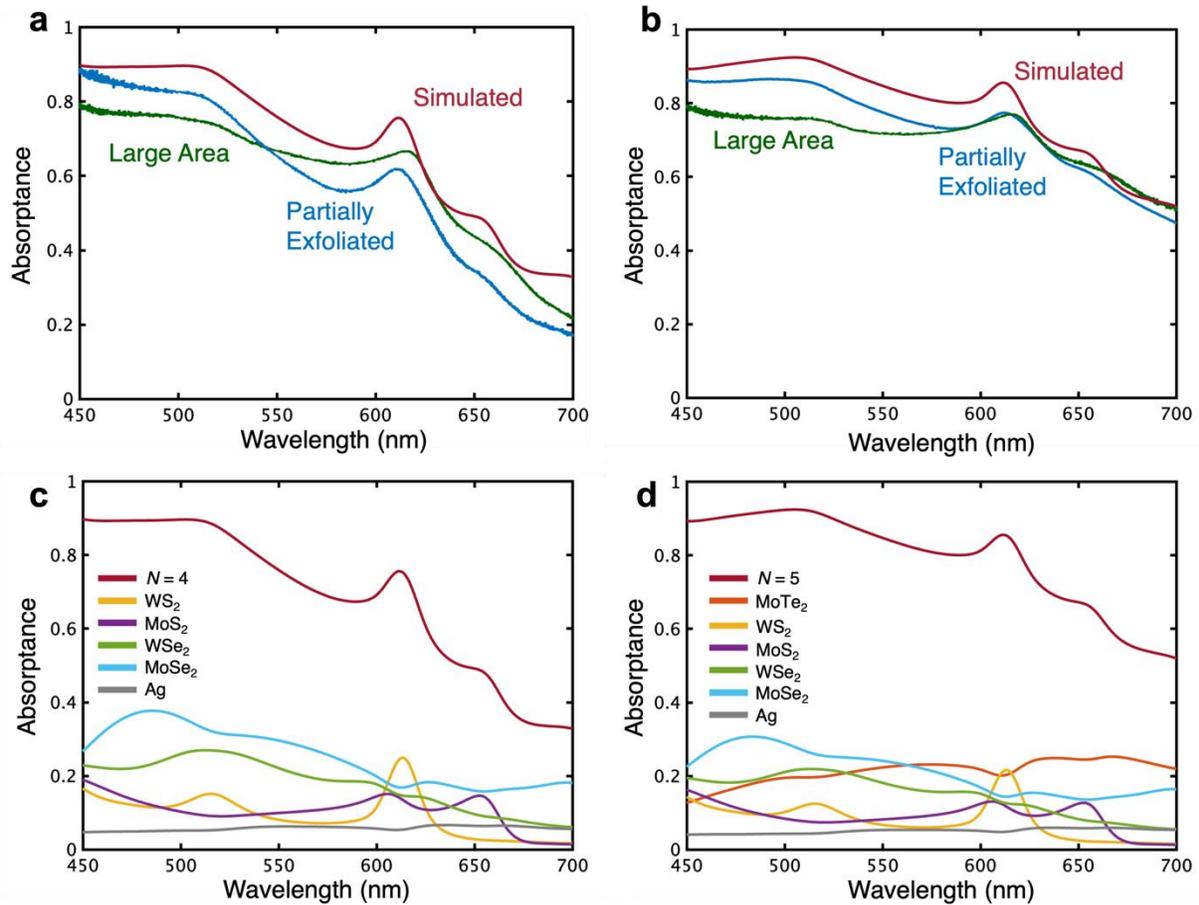

Figure 3. Measured and calculated total absorption spectra from 450 nm to 700 nm at normal incidence for the (a) $N = 4$ and (b) $N = 5$ superlattice, and the corresponding (c,d) calculated layer-by-layer total absorption spectra.

We benchmark our $N = 5$ SL against other ultrathin broadband absorbers in the field in **Figure 4** by comparing average absorption in the 450 nm – 700 nm range and effective active layer thickness. Further details about how the effective thickness and absorption is calculated is found in the **Supplementary Information**. We compare our SL to geometries ranging from plasmonic metasurfaces to planar semiconductors. Our superlattice is found to have the record average absorption in the visible range per nanometer of active layer thickness (21.1% nm$^{-1}$ for the partially-exfoliated sample and 20.1% nm$^{-1}$ for the large area sample). As seen, while various metal-based absorbers have achieved near-unity broadband absorption within the visible range, they require metals that are thicker than the penetration depth of the metals (10 to 100 nm typically). Additionally, the photocarriers in the metals lose all their energy to thermalization. Therefore, plasmonic metasurfaces are best suited for thermal applications. While plasmonic nanostructures can be used to effectively control and guide light to the underlying semiconductor layer, parasitic absorptions in the metal reduce the efficiency of the interaction while increasing the total effective thickness/weight of absorber. In addition, the need to pattern plasmonic nanostructures is a key obstacle in realizing many semiconductor-metal coupled absorbers for widespread optoelectronic use – lithographic steps are costly, timely, and technically difficult to realize and scale. Our superlattice can overcome these two key obstacles in achieving ultrathin optoelectronic devices by confining absorption almost entirely in the semiconducting layers, while employing a planar structure compatible with robust wafer-scale growth and fabrication.

While our superlattice does not have the highest visible broadband absorption of the reviewed publications, we have achieved the highest broadband absorption in a semiconductor stack < 10 nm thickness. Our device is simultaneously the thinnest by active material and most energy-dense by volume in the field, to the best of our knowledge. The reduced material usage, enhanced physical flexibility, and feasible scalability of an optimized ultrathin active absorber will open doors for an array of optoelectronic applications from photovoltaics (wearable electronics, aerospace, portable charging, etc.)[68] to sensing and stealth technology[69].

# Figure 4

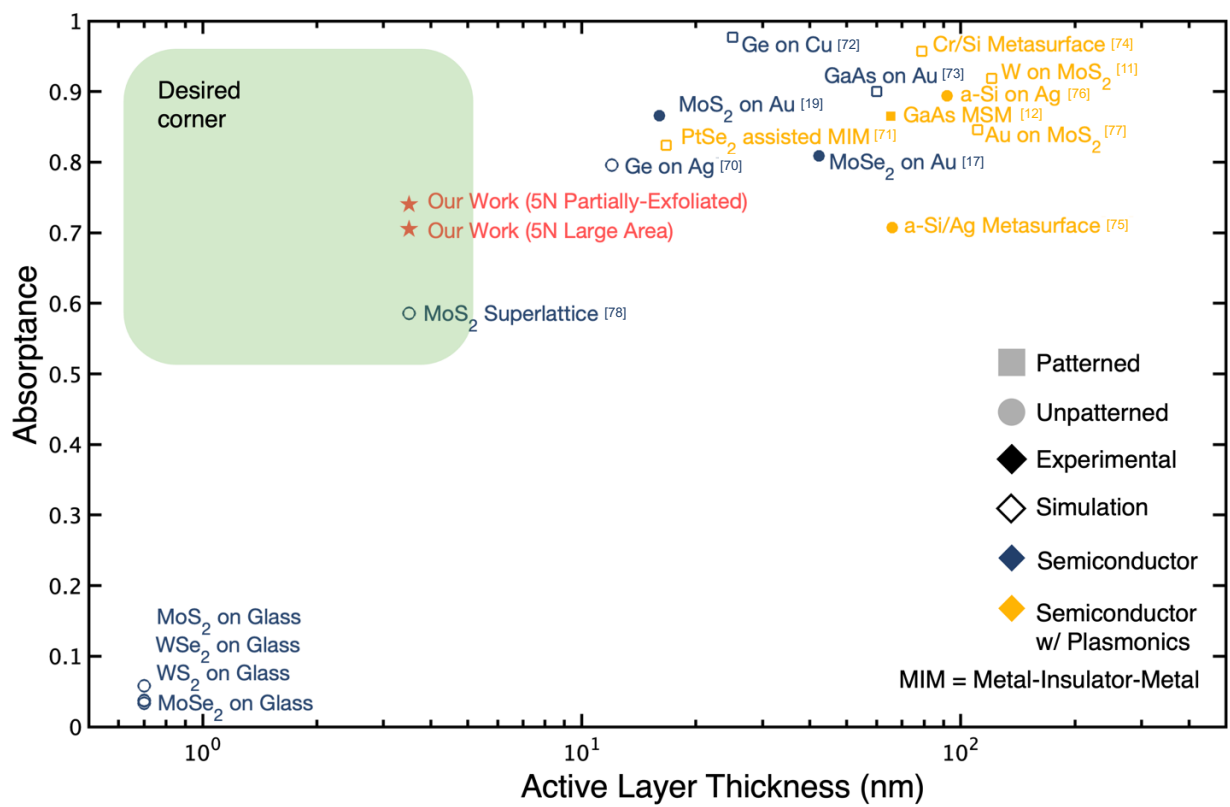

Figure 4. Benchmark of our absorber against other absorbers according to estimated absorptance and active layer thickness based on metrics of dominant absorbing material, patterning, and study type[11,12,17,19,70–78].

**Design of High Specific Power Ultrathin Photovoltaic Cell**

To elucidate the potential of our ultrathin absorber in photovoltaic (PV) applications, we consider the design of a separate contact PV cell seen in **Figure 5(a)** using a combined optoelectronic model. Silver and gold electrodes with a thickness of 0.01 $\mu$m and n- and p- regions of width 0.01 $\mu$m are used. The excitonic effects that dominate TMDC PV behaviour are accounted for as per our previous studies[78]. Experimentally demonstrated values for exciton diffusion length, radiative and non-radiative lifetimes, and binding energy of freestanding TMDC monolayers are used in the model and listed in the Supplementary Information **Table S12**. Note that binding energy is taken as 10% of the accepted exciton binding energy of a freestanding monolayer given that bandgap tuning to this extent has been demonstrated through the Coulomb engineering of the localized dielectric environment[79]. More details about the simulations conducted may be found in the Supplementary Information **Table S13**.

The performance of the ultrathin PV is optimized for **Figures 5(b),(c)**. Power conversion efficiency (PCE) scales with absorption, with the top layer MoTe$_2$ and second layer WS$_2$ being the largest contributors to J$_{SC}$ and V$_{OC}$ respectively – this is expected from the high photocurrent generated in the top layer of the SL and the large bandgap in the WS$_2$. The IV curves for the $N$ = 4 cell are provided in **Figure S14**. We thus calculate a PCE of 14.05% for the $N$ = 4 SL and 15.54% for the $N$ = 5, under AM 1.5 normal incidence. This performance can be benchmarked against other leading material classes for ultrathin PVs, such as organic PVs (OPVs) and perovskites[80,81]. In **Figure 5(d)**, the metric of specific power is used. This is calculated by dividing the converted integrated spectrum power at AM 1.5 for the device by its effective thickness, with more details on calculation provided in the **Supplementary Information**. As seen, the $N$ = 4 SL and $N$ = 5 have specific densities of 291.6 W g$^{-1}$ and 311.4 W g$^{-1}$, > 2 times higher than the next leading technology, to the best of our knowledge. This is due to the outstanding excitonic properties of TMDC-based PVs[78] and ultrathin nature of our material at 2.8 nm ($N$ = 4) and 3.5 nm ($N$ = 5) of active thickness, < 60 nm of effective thickness, and < 0.5 g m$^{-2}$ of weight.

# Figure 5

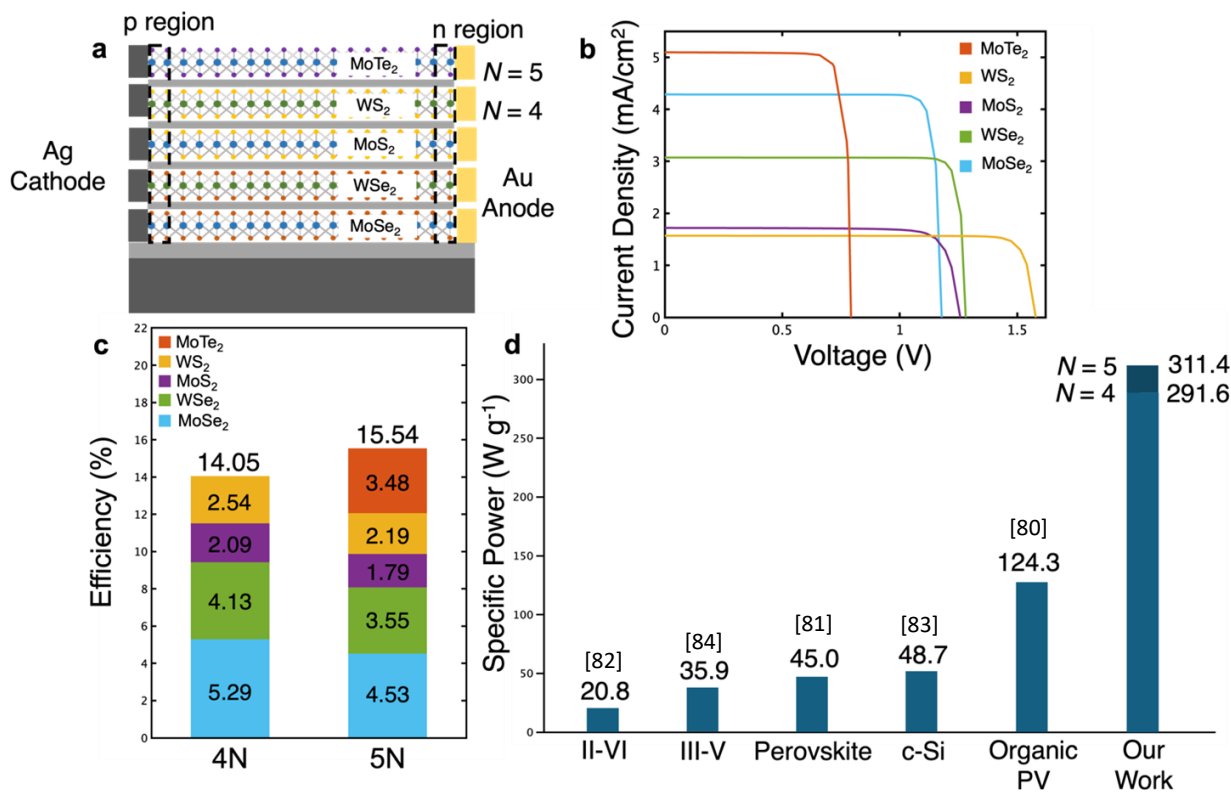

Figure 5. (a) Schematic of separate-contact model for a photovoltaic (PV) cell based on superlattice absorber. (b) Calculated I-V curves for $N$ = 5 separate contact model. (c) Layer-by-layer calculated photon conversion efficient (PCE) based on $N$ = 4 and $N$ = 5 superlattice PV. (d) Benchmark of the specific power of the $N$ = 5 superlattice PV against leading ultrathin PV materials with estimated effective thicknesses[80–84].

## **Conclusion**

In summary, we have presented a scalable TMDC-based heterostructure superlattice structure using five different monolayer TMDCs as light-trapping layers. The incorporation of the various electronically-isolated TMDCs contributes to multiple excitonic resonant peaks within the same structure, ultimately achieving > 70% average total broadband absorption in the 450 – 700 nm visible range in a superlattice using 3.5 nm of active material. Further, the superlattice is predicted to have a power conversion efficiency of 15.54% and a specific power of over 300 W g$^{-1}$ in a separate contact model optimized for excitonic effects. Our study presents a robust, scalable, and realistic approach to engineering broadband semiconductor absorbers on arbitrary surfaces with applications not only in photovoltaics but also in sensing, integrated circuits, displays, and more.

**Acknowledgements**

The authors acknowledge primary support from the Office of Naval Research (ONR) Young Investigator Award (YIP) (N00014-23-1-203) Metamaterials Program. The authors thank Adam Alfieri for his work on the alignment markers for XTEM. D.J. also acknowledges support from the Alfred P. Sloan Foundation's Sloan Fellowship in Chemistry. D.L. acknowledges support from the Penn Engineering Rachleff Scholars Program and University of Pennsylvania Center for Undergraduate Research & Fellowships Grants for Faculty Mentoring Undergraduate Research (GfFMUR). H.Z. acknowledges support from the U.S. Department of Commerce, NIST under financial assistance awards 70NANB22H101 and 70NANB22H075. The MOCVD grown TMDC monolayer samples were provided by the 2D Crystal Consortium Materials Innovation Platform (2DCC-MIP) facility at Penn State which is funded by NSF under cooperative agreement DMR-2039351.

**References**

(1) Sai, H.; Oku, T.; Sato, Y.; Tanabe, M.; Matsui, T.; Matsubara, K. Potential of Very Thin and High-efficiency Silicon Heterojunction Solar Cells. *Prog. Photovolt. Res. Appl.* **2019**, *27* (12), 1061–1070. https://doi.org/10.1002/pip.3181.

(2) Massiot, I.; Cattoni, A.; Collin, S. Progress and Prospects for Ultrathin Solar Cells. *Nat. Energy* **2020**, *5* (12), 959–972. https://doi.org/10.1038/s41560-020-00714-4.

(3) Dotan, H.; Kfir, O.; Sharlin, E.; Blank, O.; Gross, M.; Dumchin, I.; Ankonina, G.; Rothschild, A. Resonant Light Trapping in Ultrathin Films for Water Splitting. *Nat. Mater.* **2013**, *12* (2), 158–164. https://doi.org/10.1038/nmat3477.

(4) Reece, S. Y.; Hamel, J. A.; Sung, K.; Jarvi, T. D.; Esswein, A. J.; Pijpers, J. J. H.; Nocera, D. G. Wireless Solar Water Splitting Using Silicon-Based Semiconductors and Earth-Abundant Catalysts. *Science* **2011**, *334* (6056), 645–648. https://doi.org/10.1126/science.1209816.

(5) Goldflam, M. D.; Kadlec, E. A.; Olson, B. V.; Klem, J. F.; Hawkins, S. D.; Parameswaran, S.; Coon, W. T.; Keeler, G. A.; Fortune, T. R.; Tauke-Pedretti, A.; Wendt, J. R.; Shaner, E. A.; Davids, P. S.; Kim, J. K.; Peters, D. W. Enhanced Infrared Detectors Using Resonant Structures Combined with Thin Type-II Superlattice Absorbers. *Appl. Phys. Lett.* **2016**, *109* (25). https://doi.org/10.1063/1.4972844.

(6) Landy, N. I.; Sajuyigbe, S.; Mock, J. J.; Smith, D. R.; Padilla, W. J. Perfect Metamaterial Absorber. *Phys. Rev. Lett.* **2008**, *100* (20), 207402. https://doi.org/10.1103/PhysRevLett.100.207402.

(7) Aydin, K.; Ferry, V. E.; Briggs, R. M.; Atwater, H. A. Broadband Polarization-Independent Resonant Light Absorption Using Ultrathin Plasmonic Super Absorbers. *Nat. Commun.* **2011**, *2* (1), 517. https://doi.org/10.1038/ncomms1528.

(8) Li, W.; Guler, U.; Kinsey, N.; Naik, G. V.; Boltasseva, A.; Guan, J.; Shalaev, V. M.; Kildishev, A. V. Refractory Plasmonics with Titanium Nitride: Broadband Metamaterial Absorber. *Adv. Mater.* **2014**, *26* (47), 7959–7965. https://doi.org/10.1002/adma.201401874.

(9) Deng, H.; Li, Z.; Stan, L.; Rosenmann, D.; Czaplewski, D.; Gao, J.; Yang, X. Broadband Perfect Absorber Based on One Ultrathin Layer of Refractory Metal. *Opt. Lett.* **2015**, *40* (11), 2592. https://doi.org/10.1364/OL.40.002592.

(10) Lei, L.; Li, S.; Huang, H.; Tao, K.; Xu, P. Ultra-Broadband Absorber from Visible to near-Infrared Using Plasmonic Metamaterial. *Opt. Express* **2018**, *26* (5), 5686. https://doi.org/10.1364/OE.26.005686.

(11) Li, J.; Chen, X.; Yi, Z.; Yang, H.; Tang, Y.; Yi, Y.; Yao, W.; Wang, J.; Yi, Y. Broadband Solar Energy Absorber Based on Monolayer Molybdenum Disulfide Using Tungsten Elliptical Arrays. *Mater. Today Energy* **2020**, *16*, 100390. https://doi.org/10.1016/j.mtener.2020.100390.

(12) Massiot, I.; Vandamme, N.; Bardou, N.; Dupuis, C.; Lemaître, A.; Guillemoles, J.-F.; Collin, S. Metal Nanogrid for Broadband Multiresonant Light-Harvesting in Ultrathin GaAs Layers. *ACS Photonics* **2014**, *1* (9), 878–884. https://doi.org/10.1021/ph500168b.

(13) Pala, R. A.; Butun, S.; Aydin, K.; Atwater, H. A. Omnidirectional and Broadband Absorption Enhancement from Trapezoidal Mie Resonators in Semiconductor Metasurfaces. *Sci. Rep.* **2016**, *6* (1), 31451. https://doi.org/10.1038/srep31451.

(14) Tavakoli, N.; Spalding, R.; Lambertz, A.; Koppejan, P.; Gkantzounis, G.; Wan, C.; Röhrich, R.; Kontoleta, E.; Koenderink, A. F.; Sapienza, R.; Florescu, M.; Alarcon-Llado, E. Over 65% Sunlight Absorption in a 1 Mm Si Slab with Hyperuniform Texture. *ACS Photonics* **2022**, *9* (4), 1206–1217. https://doi.org/10.1021/acsphotonics.1c01668.

(15) Kats, M. A.; Blanchard, R.; Genevet, P.; Capasso, F. Nanometre Optical Coatings Based on Strong Interference Effects in Highly Absorbing Media. *Nat. Mater.* **2013**, *12* (1), 20–24. https://doi.org/10.1038/nmat3443.

(16) Alfieri, A. D.; Ruth, T.; Lim, C.; Lynch, J.; Jariwala, D. Self-Hybridized Exciton-Polariton Photovoltaics. arXiv June 18, 2024. http://arxiv.org/abs/2406.13065 (accessed 2024-06-23).

(17) Du, W.; Yu, P.; Zhu, J.; Li, C.; Xu, H.; Zou, J.; Wu, C.; Wen, Q.; Ji, H.; Liu, T.; Li, Y.; Zou, G.; Wu, J.; Wang, Z. M. An Ultrathin $MoSe_2$ Photodetector with near-Perfect Absorption. *Nanotechnology* **2020**, *31* (22), 225201. https://doi.org/10.1088/1361-6528/ab746f.

(18) Liu, D.; Yu, H.; Yang, Z.; Duan, Y. Ultrathin Planar Broadband Absorber through Effective Medium Design. *Nano Res.* **2016**, *9* (8), 2354–2363. https://doi.org/10.1007/s12274-016-1122-x.

(19) Zhang, Y.; Liu, W.; Li, Z.; Cheng, H.; Zhang, Y.; Jia, G.; Chen, S.; Tian, J. Ultrathin Polarization-Insensitive Wide-Angle Broadband near-Perfect Absorber in the Visible Regime Based on Few-Layer MoS2 Films. *Appl. Phys. Lett.* **2017**, *111* (11). https://doi.org/10.1063/1.4992045.

(20) Liu, D.; Li, Q. Sub-Nanometer Planar Solar Absorber. *Nano Energy* **2017**, *34*, 172–180. https://doi.org/10.1016/j.nanoen.2017.02.027.

(21) Pirruccio, G.; Moreno, L. M.; Lozano, G.; Rivas, J. G. Coherent and Broadband Enhanced Optical Absorption in Graphene. *ACS Nano* **2013**, *7* (6), 4810–4817. https://doi.org/10.1021/nn4012253.

(22) Kumar, P.; Lynch, J.; Song, B.; Ling, H.; Barrera, F.; Kisslinger, K.; Zhang, H.; Anantharaman, S. B.; Digani, J.; Zhu, H.; Choudhury, T. H.; McAleese, C.; Wang, X.; Conran, B. R.; Whear, O.; Motala, M. J.; Snure, M.; Muratore, C.; Redwing, J. M.; Glavin, N. R.; Stach, E. A.; Davoyan, A. R.; Jariwala, D. Light–Matter Coupling in Large-Area van Der Waals Superlattices. *Nat. Nanotechnol.* **2022**, *17* (2), 182–189. https://doi.org/10.1038/s41565-021-01023-x.

(23) Imamoto, H.; Sato, F.; Imanaka, K.; Shimura, M. AlGaAs/GaAs Superlattice Multi-Quantum-Well Laser Diode. *Superlattices Microstruct.* **1989**, *5* (2), 167–170. https://doi.org/10.1016/0749-6036(89)90276-0.

(24) Mak, K. F.; Lee, C.; Hone, J.; Shan, J.; Heinz, T. F. Atomically Thin MoS2: A New Direct-Gap Semiconductor. *Phys. Rev. Lett.* **2010**, *105* (13), 136805. https://doi.org/10.1103/PhysRevLett.105.136805.

(25) Wang, G.; Chernikov, A.; Glazov, M. M.; Heinz, T. F.; Marie, X.; Amand, T.; Urbaszek, B. *Colloquium* : Excitons in Atomically Thin Transition Metal Dichalcogenides. *Rev. Mod. Phys.* **2018**, *90* (2), 021001. https://doi.org/10.1103/RevModPhys.90.021001.

(26) Gutiérrez, H. R.; Perea-López, N.; Elías, A. L.; Berkdemir, A.; Wang, B.; Lv, R.; López-Urías, F.; Crespi, V. H.; Terrones, H.; Terrones, M. Extraordinary Room-Temperature Photoluminescence in Triangular WS$_2$ Monolayers. *Nano Lett.* **2013**, *13* (8), 3447–3454. https://doi.org/10.1021/nl3026357.

(27) Desai, S. B.; Seol, G.; Kang, J. S.; Fang, H.; Battaglia, C.; Kapadia, R.; Ager, J. W.; Guo, J.; Javey, A. Strain-Induced Indirect to Direct Bandgap Transition in Multilayer WSe$_2$. *Nano Lett.* **2014**, *14* (8), 4592–4597. https://doi.org/10.1021/nl501638a.

(28) Nie, Z.; Trovatello, C.; Pogna, E. A. A.; Conte, S. D.; Miranda, P. B.; Kelleher, E.; Zhu, C.; Turcu, I. C. E.; Xu, Y.; Liu, K.; Cerullo, G.; Wang, F. Broadband Nonlinear Optical Response of Monolayer MoSe2 under Ultrafast Excitation. *Appl. Phys. Lett.* **2018**, *112* (3). https://doi.org/10.1063/1.5010060.

(29) Tonndorf, P.; Schmidt, R.; Böttger, P.; Zhang, X.; Börner, J.; Liebig, A.; Albrecht, M.; Kloc, C.; Gordan, O.; Zahn, D. R. T.; Vasconcellos, S. M. de; Bratschitsch, R. Photoluminescence Emission and Raman Response of Monolayer MoS2, MoSe2, and WSe2. *Opt. Express* **2013**, *21* (4), 4908. https://doi.org/10.1364/OE.21.004908.

(30) Ruppert, C.; Aslan, B.; Heinz, T. F. Optical Properties and Band Gap of Single- and Few-Layer MoTe$_2$ Crystals. *Nano Lett.* **2014**, *14* (11), 6231–6236. https://doi.org/10.1021/nl502557g.

(31) Ameri, T.; Dennler, G.; Lungenschmied, C.; J. Brabec, C. Organic Tandem Solar Cells: A Review. *Energy Environ. Sci.* **2009**, *2* (4), 347–363. https://doi.org/10.1039/B817952B.

(32) Diaz, H. C.; Chaghi, R.; Ma, Y.; Batzill, M. Molecular Beam Epitaxy of the van Der Waals Heterostructure MoTe$_2$ on MoS$_2$ : Phase, Thermal, and Chemical Stability. *2D Mater.* **2015**, *2* (4), 044010. https://doi.org/10.1088/2053-1583/2/4/044010.

(33) Zhu, H.; Wang, Q.; Cheng, L.; Addou, R.; Kim, J.; Kim, M. J.; Wallace, R. M. Defects and Surface Structural Stability of MoTe$_2$ Under Vacuum Annealing. *ACS Nano* **2017**, *11* (11), 11005–11014. https://doi.org/10.1021/acsnano.7b04984.

(34) Fang, H.; Battaglia, C.; Carraro, C.; Nemsak, S.; Ozdol, B.; Kang, J. S.; Bechtel, H. A.; Desai, S. B.; Kronast, F.; Unal, A. A.; Conti, G.; Conlon, C.; Palsson, G. K.; Martin, M. C.; Minor, A. M.; Fadley, C. S.; Yablonovitch, E.; Maboudian, R.; Javey, A. Strong Interlayer Coupling in van Der Waals Heterostructures Built from Single-Layer Chalcogenides. *Proc. Natl. Acad. Sci.* **2014**, *111* (17), 6198–6202. https://doi.org/10.1073/pnas.1405435111.

(35) Xu, W.; Kozawa, D.; Zhou, Y.; Wang, Y.; Sheng, Y.; Jiang, T.; Strano, M. S.; Warner, J. H. Controlling Photoluminescence Enhancement and Energy Transfer in WS$_2$:hBN:WS$_2$ Vertical Stacks by Precise Interlayer Distances. *Small* **2020**, *16* (3). https://doi.org/10.1002/smll.201905985.

(36) Yang, J.; Yoon, A.; Lee, D.; Song, S.; Jung, I. J.; Lim, D.-H.; Jeong, H.; Lee, Z.; Lanza, M.; Kwon, S.-Y. Wafer-Scale Memristor Array Based on Aligned Grain Boundaries of 2D Molybdenum Ditelluride for Application to Artificial Synapses. *Adv. Funct. Mater.* n/a (n/a), 2309455. https://doi.org/10.1002/adfm.202309455.

(37) Holman, Z. C.; Filipič, M.; Lipovšek, B.; Wolf, S. D.; Smole, F.; Topič, M.; Ballif, C. Parasitic Absorption in the Rear Reflector of a Silicon Solar Cell: Simulation and Measurement of the Sub-Bandgap Reflectance for Common Dielectric/Metal Reflectors. *Sol. Energy Mater. Sol. Cells* **2014**, *120*, 426–430. https://doi.org/10.1016/j.solmat.2013.06.024.

(38) Huang, Y.; Pan, Y.-H.; Yang, R.; Bao, L.-H.; Meng, L.; Luo, H.-L.; Cai, Y.-Q.; Liu, G.-D.; Zhao, W.-J.; Zhou, Z.; Wu, L.-M.; Zhu, Z.-L.; Huang, M.; Liu, L.-W.; Liu, L.; Cheng, P.; Wu, K.-H.; Tian, S.-B.; Gu, C.-Z.; Shi, Y.-G.; Guo, Y.-F.; Cheng, Z. G.; Hu, J.-P.; Zhao, L.; Yang, G.-H.; Sutter, E.; Sutter, P.; Wang, Y.-L.; Ji, W.; Zhou, X.-J.; Gao, H.-J. Universal Mechanical Exfoliation of Large-Area 2D Crystals. *Nat. Commun.* **2020**, *11* (1), 2453. https://doi.org/10.1038/s41467-020-16266-w.

(39) Lu, Z.; Sun, L.; Xu, G.; Zheng, J.; Zhang, Q.; Wang, J.; Jiao, L. Universal Transfer and Stacking of Chemical Vapor Deposition Grown Two-Dimensional Atomic Layers with Water-Soluble Polymer Mediator. *ACS Nano* **2016**, *10* (5), 5237–5242. https://doi.org/10.1021/acsnano.6b00961.

(40) Li, H.; Yin, Z.; He, Q.; Li, H.; Huang, X.; Lu, G.; Fam, D. W. H.; Tok, A. I. Y.; Zhang, Q.; Zhang, H. Fabrication of Single- and Multilayer MoS$_2$ Film-Based Field-Effect Transistors for Sensing NO at Room Temperature. *Small* **2012**, *8* (1), 63–67. https://doi.org/10.1002/smll.201101016.

(41) Sebastian, A.; Pendurthi, R.; Choudhury, T. H.; Redwing, J. M.; Das, S. Benchmarking Monolayer MoS2 and WS2 Field-Effect Transistors. *Nat. Commun.* **2021**, *12* (1), 693. https://doi.org/10.1038/s41467-020-20732-w.

(42) Kim, T.; Park, H.; Joung, D.; Kim, D.; Lee, R.; Shin, C. H.; Diware, M.; Chegal, W.; Jeong, S. H.; Shin, J. C.; Park, J.; Kang, S.-W. Wafer-Scale Epitaxial 1T′, 1T′–2H Mixed, and 2H Phases MoTe2 Thin Films Grown by Metal–Organic Chemical Vapor Deposition. *Adv. Mater. Interfaces* **2018**, *5* (15), 1800439. https://doi.org/10.1002/admi.201800439.

(43) Choudhury, T. H.; Zhang, X.; Balushi, Z. Y. A.; Chubarov, M.; Redwing, J. M. Epitaxial Growth of Two-Dimensional Layered Transition Metal Dichalcogenides. *Annu. Rev. Mater. Res.*

**2020**, *50* (Volume 50, 2020), 155–177. https://doi.org/10.1146/annurev-matsci-090519-113456.

(44) Kuc, A.; Zibouche, N.; Heine, T. Influence of Quantum Confinement on the Electronic Structure of the Transition Metal Sulfide TS2. *Phys. Rev. B* **2011**, *83* (24), 245213. https://doi.org/10.1103/PhysRevB.83.245213.

(45) Yun, W. S.; Han, S. W.; Hong, S. C.; Kim, I. G.; Lee, J. D. Thickness and Strain Effects on Electronic Structures of Transition Metal Dichalcogenides: 2H-MX2 Semiconductors (M=Mo, W; X=S, Se, Te). *Phys. Rev. B* **2012**, *85* (3), 033305. https://doi.org/10.1103/PhysRevB.85.033305.

(46) Zhu, M.; Zhang, Z.; Zhang, T.; Liu, D.; Zhang, H.; Zhang, Z.; Li, Z.; Cheng, Y.; Huang, W. Exchange between Interlayer and Intralayer Exciton in $WSe_2$/$WS_2$ Heterostructure by Interlayer Coupling Engineering. *Nano Lett.* **2022**, *22* (11), 4528–4534. https://doi.org/10.1021/acs.nanolett.2c01353.

(47) Mouri, S.; Miyauchi, Y.; Matsuda, K. Tunable Photoluminescence of Monolayer $MoS_2$ via Chemical Doping. *Nano Lett.* **2013**, *13* (12), 5944–5948. https://doi.org/10.1021/nl403036h.

(48) Ghosh, C. K.; Sarkar, D.; Mitra, M. K.; Chattopadhyay, K. K. Equibiaxial Strain: Tunable Electronic Structure and Optical Properties of Bulk and Monolayer $MoSe_2$. *J. Phys. Appl. Phys.* **2013**, *46* (39), 395304. https://doi.org/10.1088/0022-3727/46/39/395304.

(49) Raja, A.; Waldecker, L.; Zipfel, J.; Cho, Y.; Brem, S.; Ziegler, J. D.; Kulig, M.; Taniguchi, T.; Watanabe, K.; Malic, E.; Heinz, T. F.; Berkelbach, T. C.; Chernikov, A. Dielectric Disorder in Two-Dimensional Materials. *Nat. Nanotechnol.* **2019**, *14* (9), 832–837. https://doi.org/10.1038/s41565-019-0520-0.

(50) Mohamed, N. B.; Lim, H. E.; Wang, F.; Koirala, S.; Mouri, S.; Shinokita, K.; Miyauchi, Y.; Matsuda, K. Long Radiative Lifetimes of Excitons in Monolayer Transition-Metal Dichalcogenides $MX_2$ ($M$ = Mo, W; $X$ = S, Se). *Appl. Phys. Express* **2018**, *11* (1), 015201. https://doi.org/10.7567/APEX.11.015201.

(51) Roy, S.; Sharbirin, A. S.; Lee, Y.; Kim, W. B.; Kim, T. S.; Cho, K.; Kang, K.; Jung, H. S.; Kim, J. Measurement of Quantum Yields of Monolayer TMDs Using Dye-Dispersed PMMA Thin Films. *Nanomaterials* **2020**, *10* (6), 1032. https://doi.org/10.3390/nano10061032.

(52) Cui, Q.; Luo, Z.; Cui, Q.; Zhu, W.; Shou, H.; Wu, C.; Liu, Z.; Lin, Y.; Zhang, P.; Wei, S.; Yang, H.; Chen, S.; Pan, A.; Song, L. Robust and High Photoluminescence in $WS_2$ Monolayer through In Situ Defect Engineering. *Adv. Funct. Mater.* **2021**, *31* (38). https://doi.org/10.1002/adfm.202105339.

(53) Wang, J.; Liu, H.; Hu, X.; Liu, Y.; Liu, D. Imaging of Defect-Accelerated Energy Transfer in $MoS_2$/hBN/$WS_2$ Heterostructures. *ACS Appl. Mater. Interfaces* **2022**, *14* (6), 8521–8526. https://doi.org/10.1021/acsami.1c20536.

(54) Li, X.; Puretzky, A. A.; Sang, X.; KC, S.; Tian, M.; Ceballos, F.; Mahjouri-Samani, M.; Wang, K.; Unocic, R. R.; Zhao, H.; Duscher, G.; Cooper, V. R.; Rouleau, C. M.; Geohegan, D. B.; Xiao, K. Suppression of Defects and Deep Levels Using Isoelectronic Tungsten Substitution in Monolayer $MoSe_2$. *Adv. Funct. Mater.* **2017**, *27* (19). https://doi.org/10.1002/adfm.201603850.

(55) Kumar, D.; Kumar, V.; Kumar, R.; Kumar, M.; Kumar, P. Electron-Phonon Coupling, Thermal Expansion Coefficient, Resonance Effect, and Phonon Dynamics in High-Quality CVD-Grown

Monolayer and Bilayer MoSe2. *Phys. Rev. B* **2022**, *105* (8), 085419. https://doi.org/10.1103/PhysRevB.105.085419.

(56) Bauccio, M. *ASM Engineered Materials Reference Book*, 2nd ed.; ASM International, 1994.

(57) Zhang, D.; Wu, Y.-C.; Yang, M.; Liu, X.; Coileáin, C. Ó.; Xu, H.; Abid, M.; Abid, M.; Wang, J.-J.; Shvets, I. V.; Liu, H.; Wang, Z.; Yin, H.; Liu, H.; Chun, B. S.; Zhang, X.; Wu, H.-C. Probing Thermal Expansion Coefficients of Monolayers Using Surface Enhanced Raman Scattering. *RSC Adv.* **2016**, *6* (101), 99053–99059. https://doi.org/10.1039/C6RA20623A.

(58) Morell, N.; Reserbat-Plantey, A.; Tsioutsios, I.; Schädler, K. G.; Dubin, F.; Koppens, F. H. L.; Bachtold, A. High Quality Factor Mechanical Resonators Based on $WSe_2$ Monolayers. *Nano Lett.* **2016**, *16* (8), 5102–5108. https://doi.org/10.1021/acs.nanolett.6b02038.

(59) Zhang, L.; Lu, Z.; Song, Y.; Zhao, L.; Bhatia, B.; Bagnall, K. R.; Wang, E. N. Thermal Expansion Coefficient of Monolayer Molybdenum Disulfide Using Micro-Raman Spectroscopy. *Nano Lett.* **2019**, *19* (7), 4745–4751. https://doi.org/10.1021/acs.nanolett.9b01829.

(60) Lee, C.; Yan, H.; Brus, L. E.; Heinz, T. F.; Hone, J.; Ryu, S. Anomalous Lattice Vibrations of Single- and Few-Layer $MoS_2$. *ACS Nano* **2010**, *4* (5), 2695–2700. https://doi.org/10.1021/nn1003937.

(61) Berkdemir, A.; Gutiérrez, H. R.; Botello-Méndez, A. R.; Perea-López, N.; Elías, A. L.; Chia, C.-I.; Wang, B.; Crespi, V. H.; López-Urías, F.; Charlier, J.-C.; Terrones, H.; Terrones, M. Identification of Individual and Few Layers of WS2 Using Raman Spectroscopy. *Sci. Rep.* **2013**, *3* (1), 1755. https://doi.org/10.1038/srep01755.

(62) Zhang, W.; Huang, J.-K.; Chen, C.-H.; Chang, Y.-H.; Cheng, Y.-J.; Li, L.-J. High-Gain Phototransistors Based on a CVD $MoS_2$ Monolayer. *Adv. Mater.* **2013**, *25* (25), 3456–3461. https://doi.org/10.1002/adma.201301244.

(63) Ye, Z.; Cao, T.; O'Brien, K.; Zhu, H.; Yin, X.; Wang, Y.; Louie, S. G.; Zhang, X. Probing Excitonic Dark States in Single-Layer Tungsten Disulphide. *Nature* **2014**, *513* (7517), 214–218. https://doi.org/10.1038/nature13734.

(64) Mattheiss, L. F. Band Structures of Transition-Metal-Dichalcogenide Layer Compounds. *Phys. Rev. B* **1973**, *8* (8), 3719–3740. https://doi.org/10.1103/PhysRevB.8.3719.

(65) Pettersson, L. A. A.; Roman, L. S.; Inganäs, O. Modeling Photocurrent Action Spectra of Photovoltaic Devices Based on Organic Thin Films. *J. Appl. Phys.* **1999**, *86* (1), 487–496. https://doi.org/10.1063/1.370757.

(66) Jariwala, D.; Davoyan, A. R.; Wong, J.; Atwater, H. A. Van Der Waals Materials for Atomically-Thin Photovoltaics: Promise and Outlook. *ACS Photonics* **2017**, *4* (12), 2962–2970. https://doi.org/10.1021/acsphotonics.7b01103.

(67) Cao, J.; Wang, J.; Yang, G.; Lu, Y.; Sun, R.; Yan, P.; Gao, S. Enhancement of Broad-Band Light Absorption in Monolayer MoS2 Using Ag Grating Hybrid with Distributed Bragg Reflector. *Superlattices Microstruct.* **2017**, *110*, 26–30. https://doi.org/10.1016/j.spmi.2017.09.008.

(68) Reese, M. O.; Glynn, S.; Kempe, M. D.; McGott, D. L.; Dabney, M. S.; Barnes, T. M.; Booth, S.; Feldman, D.; Haegel, N. M. Increasing Markets and Decreasing Package Weight for High-Specific-Power Photovoltaics. *Nat. Energy* **2018**, *3* (11), 1002–1012. https://doi.org/10.1038/s41560-018-0258-1.

(69) Yu, P.; Besteiro, L. V.; Huang, Y.; Wu, J.; Fu, L.; Tan, H. H.; Jagadish, C.; Wiederrecht, G. P.; Govorov, A. O.; Wang, Z. Broadband Metamaterial Absorbers. *Adv. Opt. Mater.* **2019**, *7* (3). https://doi.org/10.1002/adom.201800995.

(70) Park, J.; Kang, J.-H.; Vasudev, A. P.; Schoen, D. T.; Kim, H.; Hasman, E.; Brongersma, M. L. Omnidirectional Near-Unity Absorption in an Ultrathin Planar Semiconductor Layer on a Metal Substrate. *ACS Photonics* **2014**, *1* (9), 812–821. https://doi.org/10.1021/ph500093d.

(71) He, J.; Chen, C.; Liu, W.; Zhu, X.; Zheng, Y.; Wang, S.; Chen, L.; Zhang, R. Enhanced Broadband Light Absorption of Ultrathin $PtSe_2$ in Metal–Insulator–Metal Structure. *J. Phys. Appl. Phys.* **2023**, *56* (39), 395102. https://doi.org/10.1088/1361-6463/acd78e.

(72) Tang, P.; Liu, G.; Liu, X.; Fu, G.; Liu, Z.; Wang, J. Plasmonic Wavy Surface for Ultrathin Semiconductor Black Absorbers. *Opt. Express* **2020**, *28* (19), 27764. https://doi.org/10.1364/OE.402234.

(73) Huang, L.-J.; Li, J.-Q.; Lu, M.-Y.; Chen, Y.-Q.; Zhu, H.-J.; Liu, H.-Y. Broadband Visible Light Absorber Based on Ultrathin Semiconductor Nanostructures*. *Chin. Phys. B* **2020**, *29* (1), 014201. https://doi.org/10.1088/1674-1056/ab5787.

(74) Qian, Q.; Sun, T.; Yan, Y.; Wang, C. Large-Area Wide-Incident-Angle Metasurface Perfect Absorber in Total Visible Band Based on Coupled Mie Resonances. *Adv. Opt. Mater.* **2017**, *5* (13). https://doi.org/10.1002/adom.201700064.

(75) Lee, K.-T.; Ji, C.; Guo, L. J. Wide-Angle, Polarization-Independent Ultrathin Broadband Visible Absorbers. *Appl. Phys. Lett.* **2016**, *108* (3), 031107. https://doi.org/10.1063/1.4939969.

(76) Wu, S.; Ye, Y.; Luo, M.; Chen, L. Ultrathin Omnidirectional, Broadband Visible Absorbers. *J. Opt. Soc. Am. B* **2018**, *35* (8), 1825. https://doi.org/10.1364/JOSAB.35.001825.

(77) Hashemi, M.; Ansari, N.; Vazayefi, M. MoS2-Based Absorbers with Whole Visible Spectrum Coverage and High Efficiency. *Sci. Rep.* **2022**, *12* (1), 6313. https://doi.org/10.1038/s41598-022-10280-2.

(78) Hu, Z.; Lin, D.; Lynch, J.; Xu, K.; Jariwala, D. How Good Can 2D Excitonic Solar Cells Be? *Device* **2023**, *1* (1), 100003. https://doi.org/10.1016/j.device.2023.100003.

(79) Raja, A.; Chaves, A.; Yu, J.; Arefe, G.; Hill, H. M.; Rigosi, A. F.; Berkelbach, T. C.; Nagler, P.; Schüller, C.; Korn, T.; Nuckolls, C.; Hone, J.; Brus, L. E.; Heinz, T. F.; Reichman, D. R.; Chernikov, A. Coulomb Engineering of the Bandgap and Excitons in Two-Dimensional Materials. *Nat. Commun.* **2017**, *8* (1), 15251. https://doi.org/10.1038/ncomms15251.

(80) Xiong, S.; Fukuda, K.; Lee, S.; Nakano, K.; Dong, X.; Yokota, T.; Tajima, K.; Zhou, Y.; Someya, T. Ultrathin and Efficient Organic Photovoltaics with Enhanced Air Stability by Suppression of Zinc Element Diffusion. *Adv. Sci.* **2022**, *9* (8). https://doi.org/10.1002/advs.202105288.

(81) Kaltenbrunner, M.; Adam, G.; Głowacki, E. D.; Drack, M.; Schwödiauer, R.; Leonat, L.; Apaydin, D. H.; Groiss, H.; Scharber, M. C.; White, M. S.; Sariftci, N. S.; Bauer, S. Flexible High Power-per-Weight Perovskite Solar Cells with Chromium Oxide–Metal Contacts for Improved Stability in Air. *Nat. Mater.* **2015**, *14* (10), 1032–1039. https://doi.org/10.1038/nmat4388.

(82) Gupta, A.; Parikh, V.; Compaan, A. D. High Efficiency Ultra-Thin Sputtered CdTe Solar Cells. *Sol. Energy Mater. Sol. Cells* **2006**, *90* (15), 2263–2271. https://doi.org/10.1016/j.solmat.2006.02.029.

(83) Zhao, F.; Lin, J.; Lei, Z.; Yi, Z.; Qin, F.; Zhang, J.; Liu, L.; Wu, X.; Yang, W.; Wu, P. Realization of 18.97% Theoretical Efficiency of 0.9 Mm Thick c-Si/ZnO Heterojunction Ultrathin-Film Solar Cells *via* Surface Plasmon Resonance Enhancement. *Phys. Chem. Chem. Phys.* **2022**, *24* (8), 4871–4880. https://doi.org/10.1039/D1CP05119A.

(84) Buencuerpo, J.; Saenz, T. E.; Steger, M.; Young, M.; Warren, E. L.; Geisz, J. F.; Steiner, M. A.; Tamboli, A. C. Efficient Light-Trapping in Ultrathin GaAs Solar Cells Using Quasi-Random Photonic Crystals. *Nano Energy* **2022**, *96*, 107080. https://doi.org/10.1016/j.nanoen.2022.107080.23

# Supporting Information:
# Broadband Light Harvesting from Scalable Two-Dimensional Semiconductor Heterostructures

Da Lin[1], Jason Lynch[2], Sudong Wang[2], Zekun Hu[2], Rajeev Kumar Rai[1], Huairuo Zhang[3,4], Chen Chen[5], Shalini Kumari[5], Eric Stach[1], Albert V. Davydov[3], Joan M. Redwing[5, 6], Deep Jariwala[1,2]

[1] Materials Science and Engineering, University of Pennsylvania, Philadelphia, PA, USA

[2] Electrical and Systems Engineering, University of Pennsylvania, Philadelphia, PA, USA

[3] Materials Science and Engineering Division, National Institute of Standards and Technology, Gaithersburg, MD, USA

[4] Theiss Research, Inc., La Jolla, CA 92037, USA

[5] 2D Crystal Consortium Materials Innovation Platform, Materials Research Institute, Penn State University, University Park, PA, USA

[6] Materials Science and Engineering, Penn State University, University Park, PA, USA

**Experimental Procedure**

**A. Theoretical Simulations of Optical Performance**

We converted an open-source transfer matrix method[1] from MATLAB to Python and modify the source code to be suitable for heterostructure optimization. All refractive indices used for the monolayer TMDCs, dielectrics, and Ag were measured experimentally with large-area samples using ellipsometry. Absorption (1 – Reflection) was studied in terms of TMDC ordering, interlayer dielectric thicknesses, and bottom dielectric thickness, wherein the bottom dielectric thickness was ultimately determined to be the only engineerable variable of the heterostructure. A sequential least squares programming (SLSQP) algorithm was used to find the optimized bottom dielectric thickness by minimizing the negative of the total absorption calculated from 400 nm – 700 nm. A J.A. Woollam ellipsometer was used to acquire all the necessary refractive indices for the optical simulations from samples grown by MOCVD[2].

**B. Heterostructure Fabrication**

All samples are fabricated on 100 nm $SiO_2$/Si wafer, with 200 nm of Ag deposited on top via e-beam evaporation using an instrument from K.J. Lasker. Atomic layer deposition was used to deposit $Al_2O_3$ layers wherein a metal precursor and water vapor is cycled. All wafer-scale TMDCs used (monolayer $WS_2$, $MoS_2$, $WSe_2$, $MoSe_2$) are grown on sapphire substrates via MOCVD and have been previously detailed in other studies[2–5]. Large area monolayer TMDCs are released from sapphire substrates with KOH solution prior to being transferred to the target substrate and rinsed with acetone and IPA. All mechanically-exfoliated samples were prepared from bulk crystals acquired from hq-Graphene using the scotch tape method and then rinsed with acetone and IPA.

## C. Heterostructure Characterization

The PL and Raman spectra were acquired of the heterostructure throughout the fabrication process using the integrated system available with the Horiba Scientific confocal microscopy (LabRAM HR Evolution). Continuous-wave lasers of 405 nm, 532 nm, and 633 nm were used to acquire the PL and Raman, using x100 and x50 objective lenses. Temperature-dependant PL studies were conducted using the same setup with the support of a Linkam cooling stage. The Raman spectra is calibrated against a standard crystalline silicon wafer with a Raman shift at 520 $cm^{-1}$. The reflectance spectrum is also collected using the confocal system with a visible white light irradiation source provided through a fibre probe. The collected reflectance spectrum is normalized against background light and the total incident light, as measured using an Ag mirror.

## D. Solar Cell Simulations

The 2D solid-state p-i-n superlattice structure was simulated using Synopsys' Sentaurus technology CAD tools, allowing a comprehensive analysis of solar cell behavior. The simulation used a default 1 mm thickness in the third dimension, with minimal anticipated impact on results. The structure included monolayer active materials ($MoS_2$, $MoSe_2$, $WS_2$, $WSe_2$, $MoTe_2$), insulators ($Al_2O_3$), cathode (Ag), and anode (Au). Sentaurus excluded the bottom insulator and gold layer, accounting for their effects elsewhere. The study investigated the influence of various parameters (binding energy, electron diffusion length (EDL), exciton radiative/nonradiative lifetimes, carrier mobility, Shockley-Read-Hall (SRH) lifetime, and device length) on open-circuit voltage ($V_{OC}$), short-circuit current ($J_{SC}$), fill factor (FF), and power conversion efficiency (PCE). Initial parameters were derived from experimental and theoretical data to define materials and structure. The model computed electron and hole densities based on quasi-Fermi potentials and considered carrier recombination (SRH, Auger, radiative). It also permitted discontinuous interfaces in the superlattice structure and manually defined optical generation rates for each layer, with Sentaurus solving Poisson and continuity equations for optical properties. The optical generation results were obtained with TMM Python calculation and were defined manually for each layer. The carrier recombination contains three major types, SRH, Auger, and Radiative recombination. The optical properties Sentaurus solved are the results of the Poisson and continuity equation. The equation modeled the dynamic of the generation, diffusion, recombination, and radiative decay of singlet excitons is given by[6,7]:

$$\frac{\partial n_{se}}{\partial t} = R_{biomolec} + \nabla * D_{se}\nabla n_{se} - \frac{n_{se} - n_{se}^{eq}}{\tau} - \frac{n_{se} - n_{se}^{eq}}{\tau_{trap}} - R_{se}$$

Where $n_{se}$ is the singlet exciton density, $R_{biomolec}$ is the carrier bimolecular recombination rate acting as a singlet exciton generation term, $D_{se}$ is the singlet exciton diffusion constant, $\tau$, $\tau_{trap}$ are the singlet exciton lifetimes. $R_{se}$ is the net singlet exciton recombination rate.

### E. Effective Active Layer and Absorption Estimations

The active layer is evaluated individually for every study reviewed in this letter. As there lacks a uniform definition for what constitutes an "active layer", it is defined in this study as the thickness of material that interacts with light and thus experiences energy transfer (absorption) in a useful way (metals or semiconductors). This is assessed using electric field simulation figures, when provided in the reviewed study, by primarily accounting for the thicknesses in which the electric field interacts with the material. In our approach, planar back reflectors for semiconducting absorbers, such as that of Au underneath planar Ge[8], are not included in the active thickness estimation if the metal layer experiences minimal interactions with the incident EM wave. Alternatively, if significant interactions are observed, such as in one study of planar MoSe$_2$ on Au[9], an effective thickness for the metal is accounted for by the penetration depth of light into the material at 500 nm. For plasmonic absorbers, all metal nanostructures are included in the thickness estimation given their integral role in inducing surface plasma polaritons in the underlying layer. Similarly, metal-insulator-metal (MIM) active layers are taken as the full thickness of the top metal and the effective penetration depth at 500 nm of the bottom metal given that Fabry-Perot cavities function via the interaction of two reflectors[10].

The absorption spectrum for a given study is obtained for the 450 nm – 700 nm range by estimating various datapoints for different wavelengths along the provided absorption spectra and averaging the sum to obtain an estimated total absorption.

**F. Specific Power Calculations**

The approach to calculating specific power is employed directly from our previous work[11]. It obtained by dividing the total converted power by the effective weight of the device. Total converted power (W $m^{-2}$) is calculated by multiplying the PCE obtained from the study by the integrated power of the AM 1.5 spectrum (1000 W $m^{-2}$). The effective weight is defined to be the minimum device weight (g $m^{-2}$) required to produce a functioning device. It accounts for all active portions (semiconductor, electrodes, etc) and their areal coverage within the device. For components of the device that don't directly contribute to power conversion but affect the photocarrier generation rate (dielectrics, back reflector, etc.), an effective thickness is estimated from the penetration depth of light into the material at 500 nm. Supporting substrates are not included in the effective weight estimation given that they do not affect electrical nor optical behaviour and are arbitrarily chosen from study to study.

**Absorption spectra for *N* = 5 SL with different MoTe$_2$ placement**

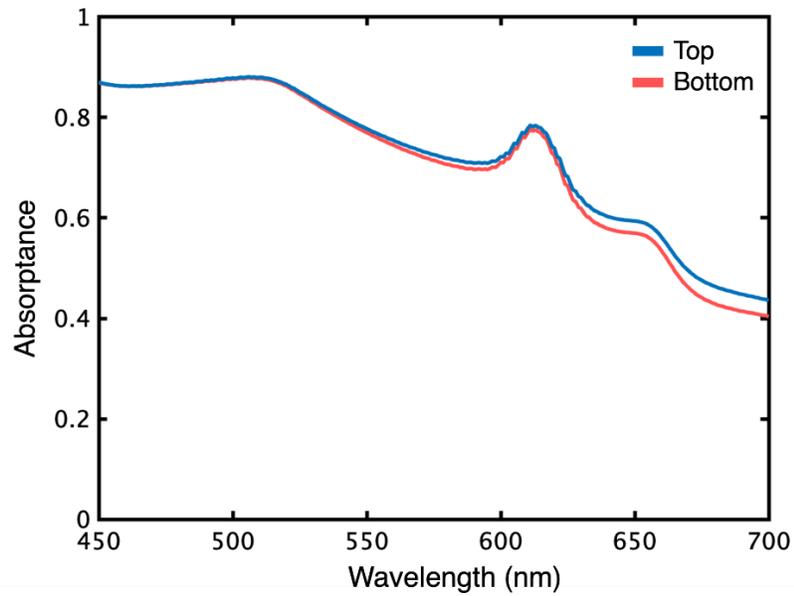

Figure S1. Absorption spectra for *N* = 5 SL with MoTe$_2$ on the top of the SL (above WS$_2$) and below the SL (below MoSe$_2$) from 450 nm – 700 nm as simulated through the TMM method.

**Optical constants for simulated materials**

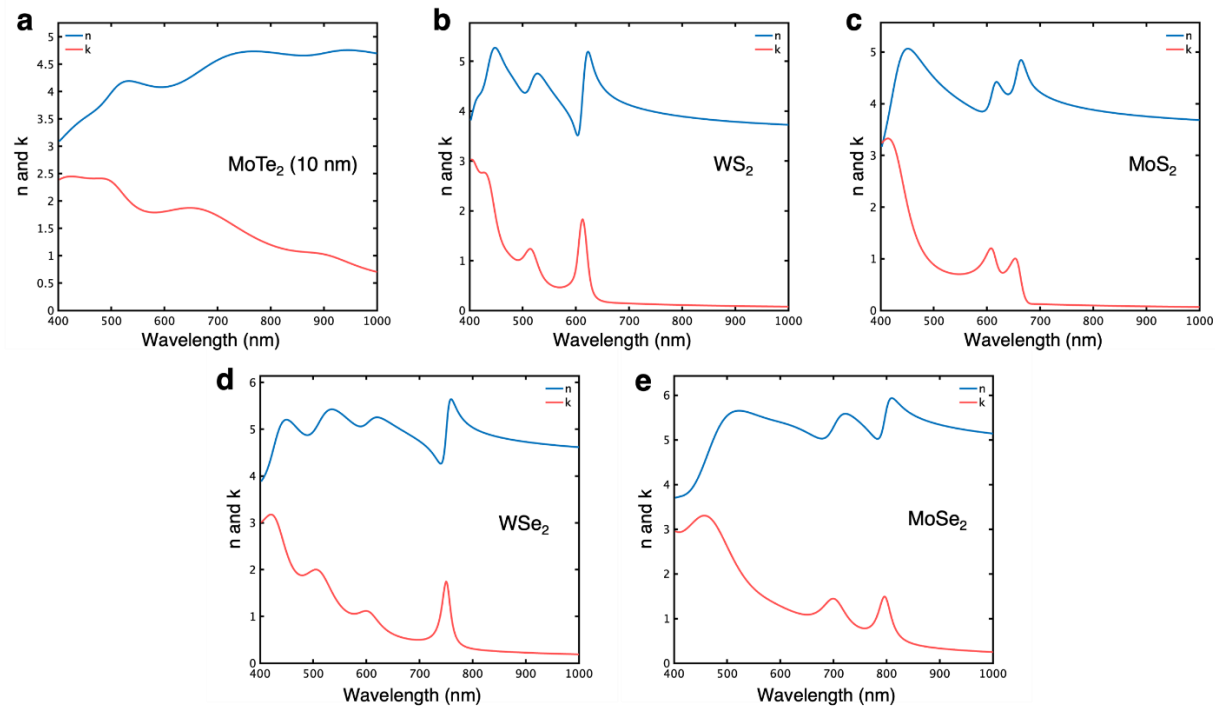

Figure S2. Measured optical constants for (a) 10 nm $MoTe_2$, (b) monolayer $WS_2$, (c) monolayer $MoS_2$, (d) monolayer $WSe_2$, (e) monolayer $MoSe_2$ grown via CVD and acquired through ellipsometry as described in the experimental procedure.

**Experimental Realization of Partially Exfoliated *N* = 5 SL**

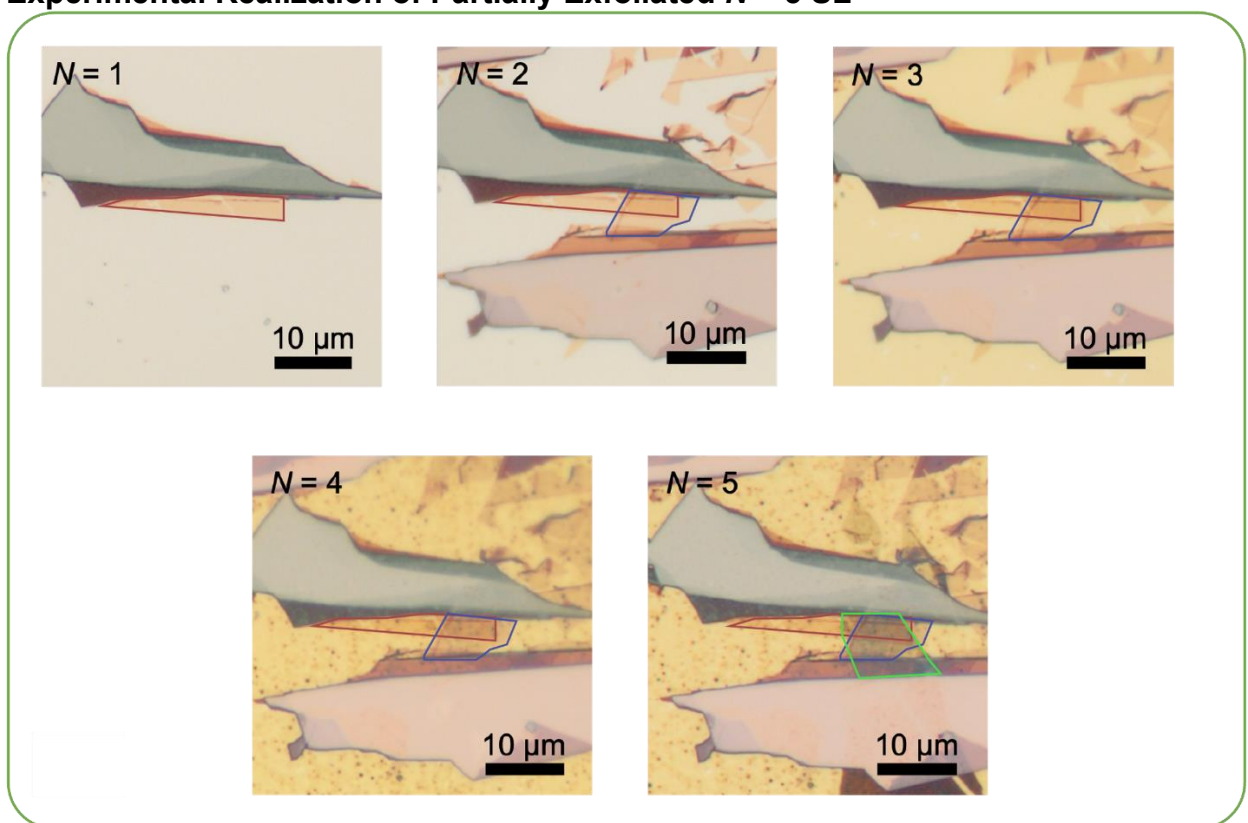

Figure S3. Optical microscope (OM) images of the step-by-step fabrication process of a *N* = 5 partially exfoliated sample device.

**TEM and EDS Analysis of Large Area *N* = 4 Superlattice Stack**

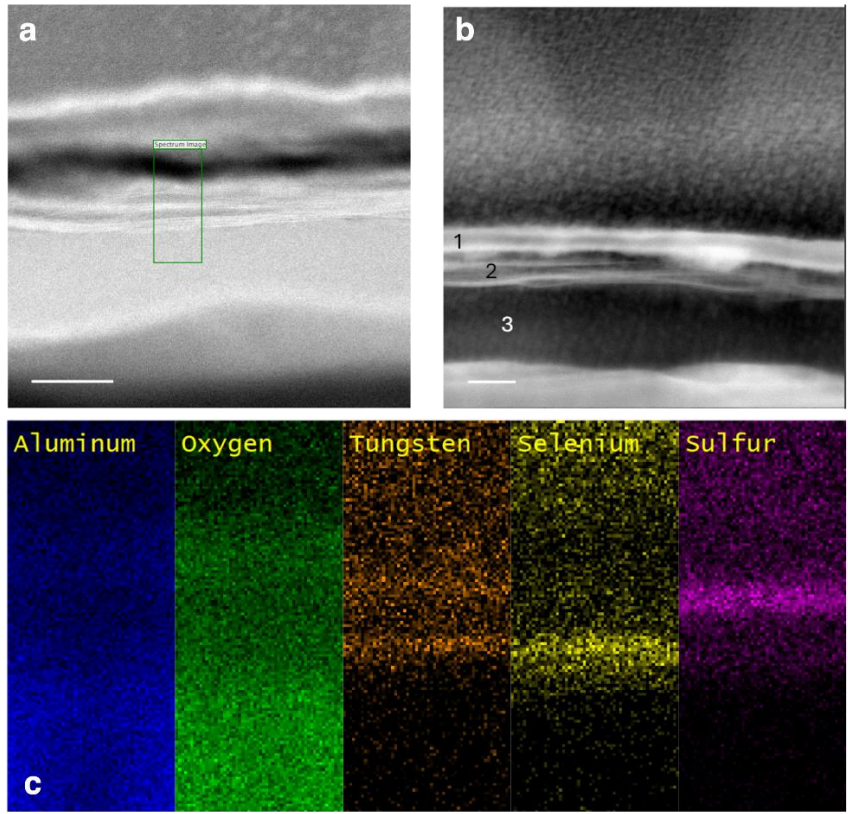

Figure S4. (a) and (b): Transmission electron microscope (TEM) images of *N* = 4 large area superlattice stack composed of monolayer CVD-grown wafer-scale TMDCs, where (b) magnified the selected area in (a) Region 1 is the sulfide layers, region 2 is the selenide layers, and region three is the bottom $Al_2O_3$ dielectric. (c): Energy dispersive x-ray spectroscopy (EDS) mapping conducted on the target area shows signs of respective layer material identities.

**PL comparison through fabrication process for *N* = 5 partially exfoliated sample**

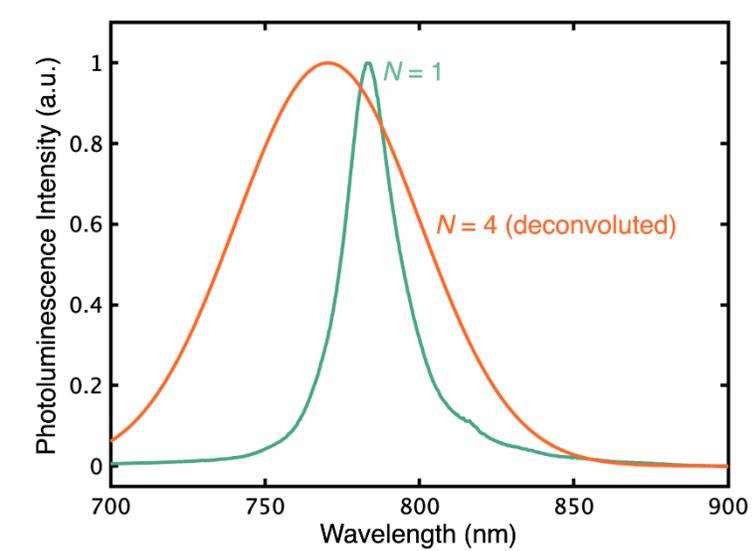

Figure S5. Normalized room temperature photoluminescence (PL) spectrum for *N* = 1 (MoSe$_2$) step of SL fabrication and the deconvoluted MoSe$_2$ peak for the *N* = 4 SL.

**Relative absorption of TMDCs in SL**

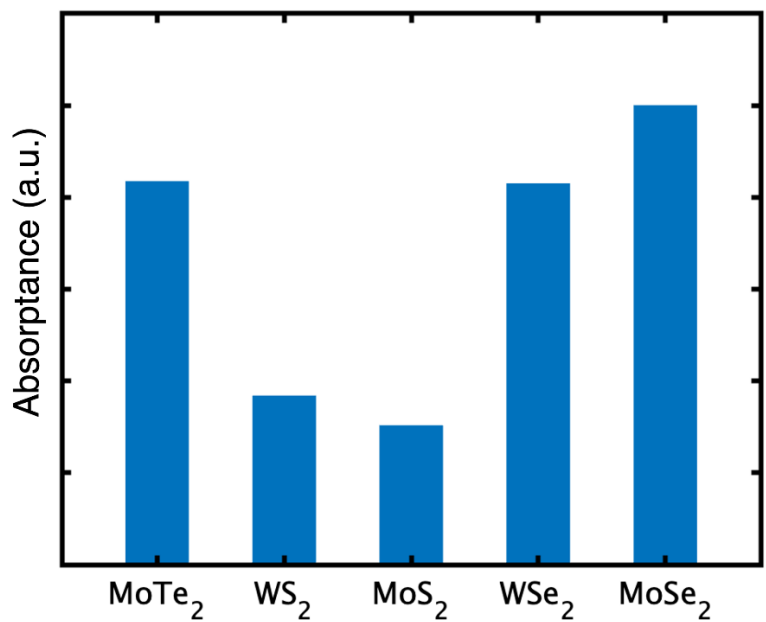

Figure S6. Relative absorption of individual layers in the *N* = 5 SL at 532 nm as simulated through the TMM method.

**Deconvolution of 80 K PL spectra for *N* = 5 partially exfoliated sample**

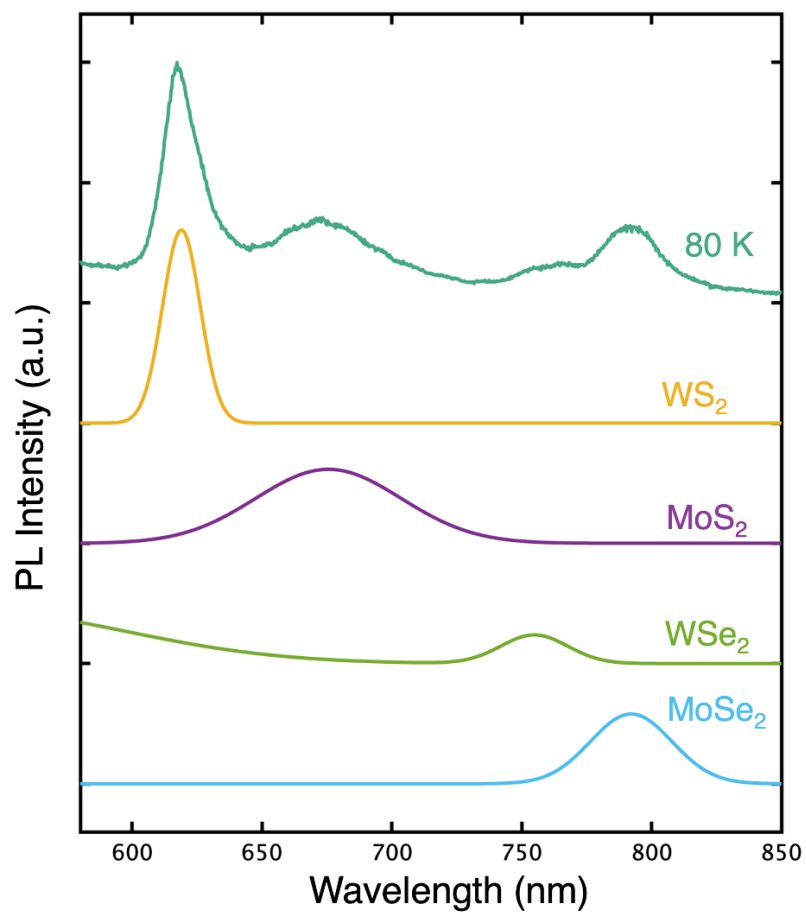

Figure S7. Gaussian deconvolution of individual TMDC PL peak contributions to the 80 K PL spectra for *N* = 5 SL given in Figure 2(a).

**Layer-by-layer Raman spectra for *N* = 3 SL partially exfoliated sample**

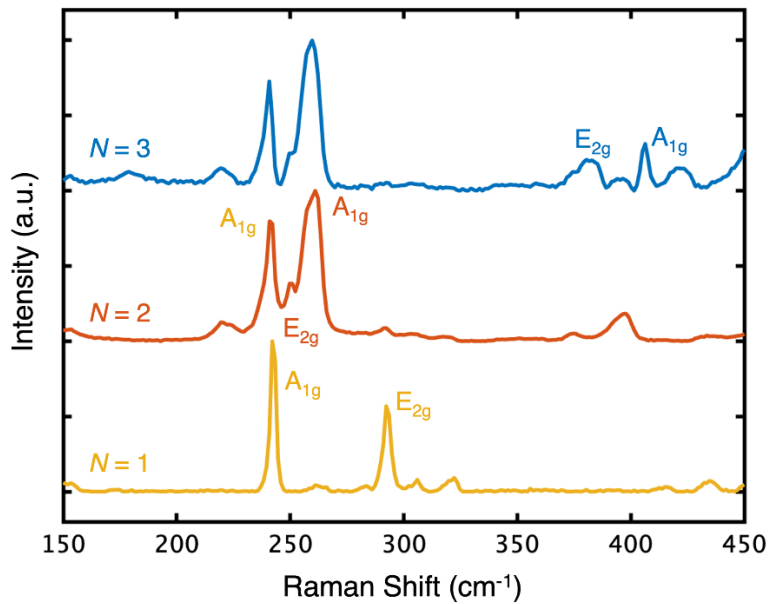

Figure S8. Layer-by-layer Raman spectra for a partially-exfoliated heterostructure building up to three layers, consisting, bottom-to-top, of $MoSe_2$, $WSe_2$, and $MoS_2$. The individual $A_{1g}$ and $E_{2g}$ peaks for each monolayer TMDC can be identified within the subsequent layer following the initial deposition of the monolayer TMDC[12].

**Photoluminescence (PL) and Raman measurements for *N* = 5 large area sample**

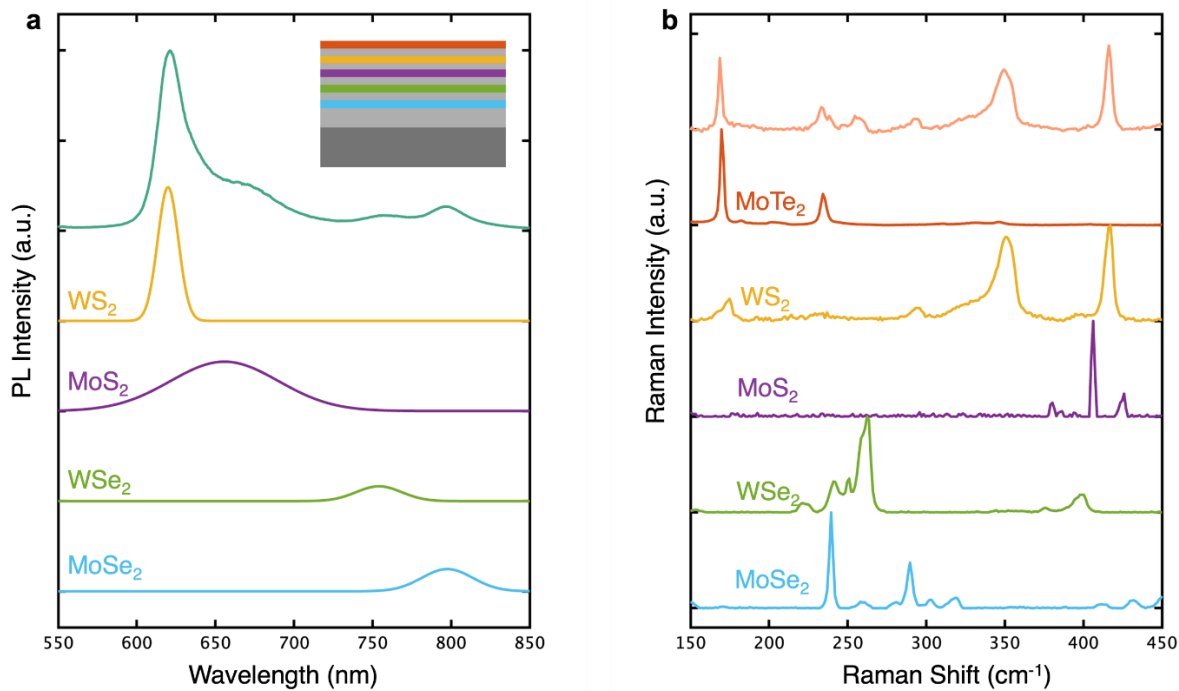

Figure S9. (a) Normalized photoluminescence (PL) spectra for the superlattice and deconvoluted Gaussian fits for individual layer PL spectra for the large area superlattice. (b) Normalized Raman spectra for superlattice matched with experimental Raman spectra for CVD-grown/dry-transferred TMDC monolayer.

**Extended absorptance spectra for *N* = 5 SL**

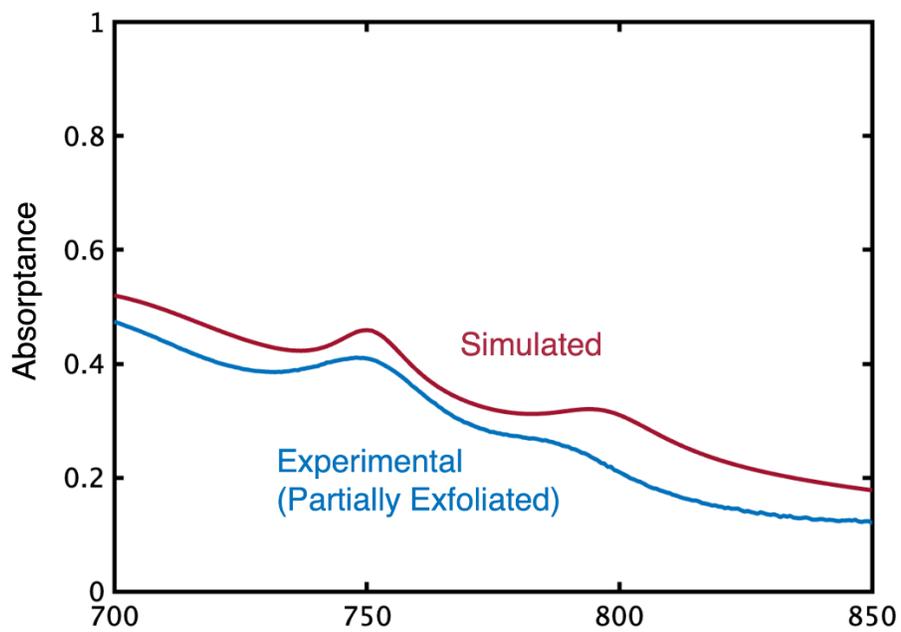

Figure S10. Measured and calculated total absorptance of the *N* = 5 partially exfoliated SL from 700 nm to 850 nm at normal incidence.

**Solar absorptance of *N* = 4 and *N* = 5 SL**

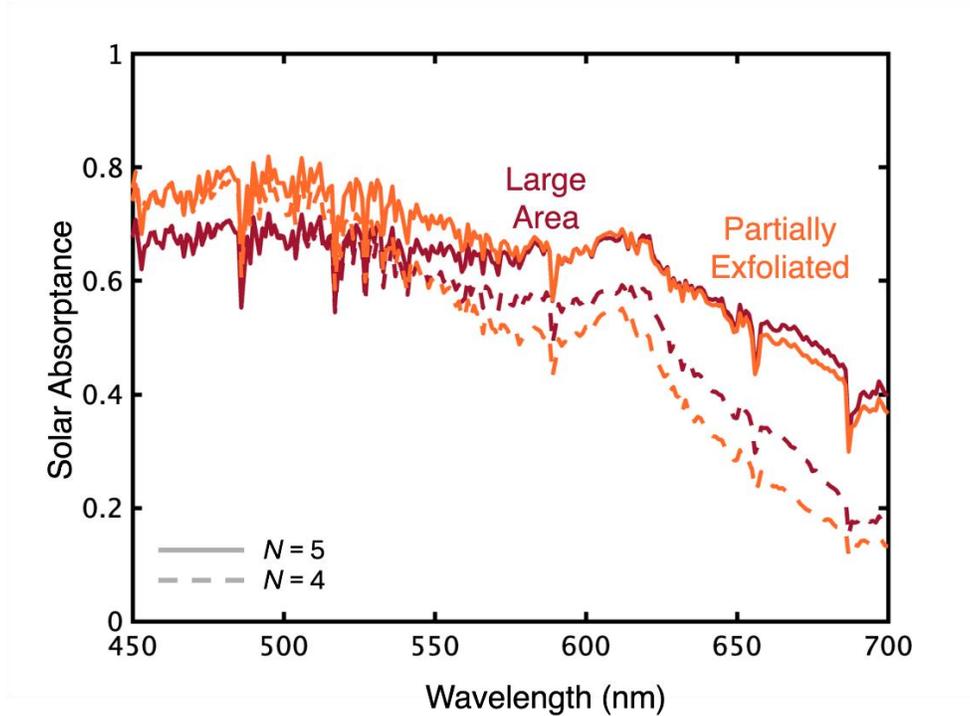

Figure S11. Absorptance of *N* = 4 and *N* = 5 partially exfoliated and large area SL in the 450 nm – 700 nm range calculated from the total collected photons within the AM 1.5 spectrum at the equator. For the large area sample, the *N* = 5 SL collects 70.8% of total incident photons within the spectrum, and the *N* = 4 SL collects 61.2%. For the partially exfoliated sample, the *N* = 5 SL collects 74.2% of total incident photons within the spectrum, and the *N* = 4 SL collects 59.4%.

Table S12. Shared values used within the simulations.

| | |
|---|---|
| Acceptor Concentration ($cm^{-3}$) | 1.00E-19 |
| Donor Concentration ($cm^{-3}$) | 1.00E-19 |
| Temperature (K) | 300 |
| Electron Lifetime (s) | 1.50E-19 |
| Hole Lifetime (s) | 1.50E-19 |

Table S13. Excitonic material parameters used for the simulations, wherein the electric bandgap is accounted for with 10% binding energy[13–25].

| | $MoS_2$ | $MoSe_2$ | $WS_2$ | $WSe_2$ | $MoTe_2$ |
|---|---|---|---|---|---|
| Bandgap (eV) | 1.80 | 1.60 | 2.04 | 1.65 | 1.19 |
| Binding Energy (eV) | 0.48 | 0.57 | 0.32 | 0.37 | 0.19 |
| Exciton Diffusion Length (μm) | 1.50 | 0.40 | 0.35 | 0.16 | 0.35 |
| Radiative Exciton Lifetime (ns) | 8.00 | 0.80 | 4.40 | 3.50 | 0.95 |
| Free Carrier Lifetime (ns) | 10 | 130 | 22 | 18 | 0.06 |
| Free Carrier Mobility ($cm^2/V*s$) | 60 | 480 | 1060 | 250 | 200 |

**IV curve for *N* = 4 separate contact solar cell**

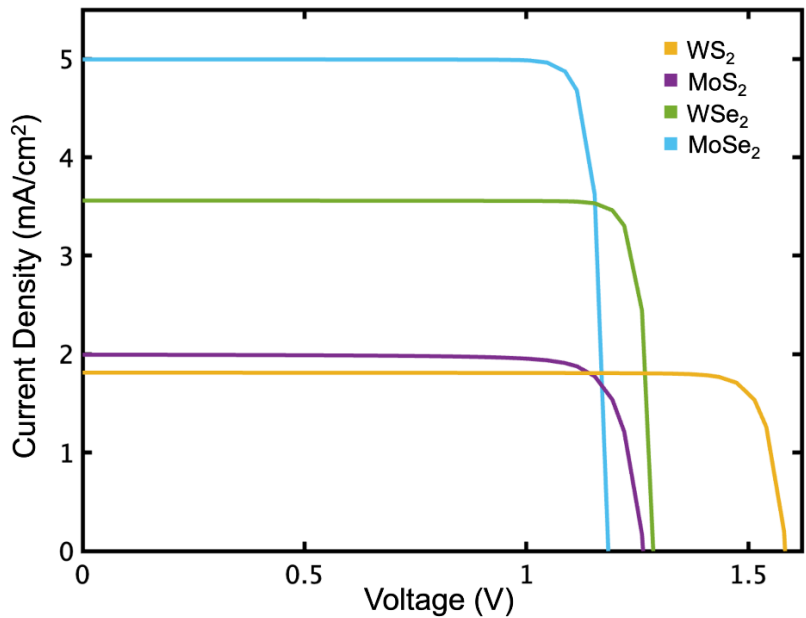

Figure S14. Calculated I-V curves for *N* = 4 separate contact model using the materials and model parameters described in Table S9 and Table S10.

Disclaimer: Certain commercial equipment, instruments, software, or materials are identified in this paper in order to specify the experimental procedure adequately. Such identifications are not intended to imply recommendation or endorsement by NIST, nor it is intended to imply that the materials or equipment identified are necessarily the best available for the purpose.

**References**

(1) Burkhard, G. F.; Hoke, E. T. Transfer Matrix Optical Modeling, 2011.
(2) Choudhury, T. H.; Zhang, X.; Balushi, Z. Y. A.; Chubarov, M.; Redwing, J. M. Epitaxial Growth of Two-Dimensional Layered Transition Metal Dichalcogenides. *Annu. Rev. Mater. Res.* **2020**, *50* (Volume 50, 2020), 155–177. https://doi.org/10.1146/annurev-matsci-090519-113456.
(3) Eichfeld, S. M.; Hossain, L.; Lin, Y.-C.; Piasecki, A. F.; Kupp, B.; Birdwell, A. G.; Burke, R. A.; Lu, N.; Peng, X.; Li, J.; Azcatl, A.; McDonnell, S.; Wallace, R. M.; Kim, M. J.; Mayer, T. S.; Redwing, J. M.; Robinson, J. A. Highly Scalable, Atomically Thin WSe2 Grown via Metal–Organic Chemical Vapor Deposition. *ACS Nano* **2015**, *9* (2), 2080–2087. https://doi.org/10.1021/nn5073286.
(4) Chubarov, M.; Choudhury, T. H.; Hickey, D. R.; Bachu, S.; Zhang, T.; Sebastian, A.; Bansal, A.; Zhu, H.; Trainor, N.; Das, S.; Terrones, M.; Alem, N.; Redwing, J. M. Wafer-Scale Epitaxial Growth of Unidirectional WS2 Monolayers on Sapphire. *ACS Nano* **2021**, *15* (2), 2532–2541. https://doi.org/10.1021/acsnano.0c06750.
(5) Sebastian, A.; Pendurthi, R.; Choudhury, T. H.; Redwing, J. M.; Das, S. Benchmarking Monolayer MoS2 and WS2 Field-Effect Transistors. *Nat. Commun.* **2021**, *12* (1), 693. https://doi.org/10.1038/s41467-020-20732-w.
(6) Ruhstaller, B.; Carter, S. A.; Barth, S.; Riel, H.; Riess, W.; Scott, J. C. Transient and Steady-State Behavior of Space Charges in Multilayer Organic Light-Emitting Diodes. *J. Appl. Phys.* **2001**, *89* (8), 4575–4586. https://doi.org/10.1063/1.1352027.
(7) Ruhstaller, B.; Beierlein, T.; Riel, H.; Karg, S.; Scott, J. C.; Riess, W. Simulating Electronic and Optical Processes in Multilayer Organic Light-Emitting Devices. *IEEE J. Sel. Top. Quantum Electron.* **2003**, *9* (3), 723–731. https://doi.org/10.1109/JSTQE.2003.818852.
(8) Park, J.; Kang, J.-H.; Vasudev, A. P.; Schoen, D. T.; Kim, H.; Hasman, E.; Brongersma, M. L. Omnidirectional Near-Unity Absorption in an Ultrathin Planar Semiconductor Layer on a Metal Substrate. *ACS Photonics* **2014**, *1* (9), 812–821. https://doi.org/10.1021/ph500093d.
(9) Du, W.; Yu, P.; Zhu, J.; Li, C.; Xu, H.; Zou, J.; Wu, C.; Wen, Q.; Ji, H.; Liu, T.; Li, Y.; Zou, G.; Wu, J.; Wang, Z. M. An Ultrathin MoSe$_2$ Photodetector with near-Perfect Absorption. *Nanotechnology* **2020**, *31* (22), 225201. https://doi.org/10.1088/1361-6528/ab746f.
(10) Kasap, S. *Optoelectronics and Photonics: Principles and Practices*; Pearson Education, 2013.
(11) Hu, Z.; Lin, D.; Lynch, J.; Xu, K.; Jariwala, D. How Good Can 2D Excitonic Solar Cells Be? *Device* **2023**, *1* (1), 100003. https://doi.org/10.1016/j.device.2023.100003.
(12) Tonndorf, P.; Schmidt, R.; Böttger, P.; Zhang, X.; Börner, J.; Liebig, A.; Albrecht, M.; Kloc, C.; Gordan, O.; Zahn, D. R. T.; Vasconcellos, S. M. de; Bratschitsch, R. Photoluminescence Emission and Raman Response of Monolayer MoS_2, MoSe_2, and WSe_2. *Opt. Express* **2013**, *21* (4), 4908. https://doi.org/10.1364/OE.21.004908.
(13) Hill, H. M.; Rigosi, A. F.; Roquelet, C.; Chernikov, A.; Berkelbach, T. C.; Reichman, D. R.; Hybertsen, M. S.; Brus, L. E.; Heinz, T. F. Observation of Excitonic Rydberg States in Monolayer MoS2 and WS2 by Photoluminescence Excitation

Spectroscopy. *Nano Lett.* **2015**, *15* (5), 2992–2997. https://doi.org/10.1021/nl504868p.

(14) Ugeda, M. M.; Bradley, A. J.; Shi, S.-F.; da Jornada, F. H.; Zhang, Y.; Qiu, D. Y.; Ruan, W.; Mo, S.-K.; Hussain, Z.; Shen, Z.-X.; Wang, F.; Louie, S. G.; Crommie, M. F. Giant Bandgap Renormalization and Excitonic Effects in a Monolayer Transition Metal Dichalcogenide Semiconductor. *Nat. Mater.* **2014**, *13* (12), 1091–1095. https://doi.org/10.1038/nmat4061.

(15) He, K.; Kumar, N.; Zhao, L.; Wang, Z.; Mak, K. F.; Zhao, H.; Shan, J. Tightly Bound Excitons in Monolayer WSe2. *Phys. Rev. Lett.* **2014**, *113* (2), 026803. https://doi.org/10.1103/PhysRevLett.113.026803.

(16) Li, J.-H.; Bing, D.; Wu, Z.-T.; Wu, G.-Q.; Bai, J.; Du, R.-X.; Qi, Z.-Q. Thickness-Dependent Excitonic Properties of Atomically Thin 2H-MoTe2*. *Chin. Phys. B* **2020**, *29* (1), 017802. https://doi.org/10.1088/1674-1056/ab5a3a.

(17) Uddin, S. Z.; Kim, H.; Lorenzon, M.; Yeh, M.; Lien, D.-H.; Barnard, E. S.; Htoon, H.; Weber-Bargioni, A.; Javey, A. Neutral Exciton Diffusion in Monolayer MoS2. *ACS Nano* **2020**, *14* (10), 13433–13440. https://doi.org/10.1021/acsnano.0c05305.

(18) Kumar, N.; Cui, Q.; Ceballos, F.; He, D.; Wang, Y.; Zhao, H. Exciton Diffusion in Monolayer and Bulk MoSe2. *Nanoscale* **2014**, *6* (9), 4915–4919. https://doi.org/10.1039/C3NR06863C.

(19) Cui, Q.; Ceballos, F.; Kumar, N.; Zhao, H. Transient Absorption Microscopy of Monolayer and Bulk WSe2. *ACS Nano* **2014**, *8* (3), 2970–2976. https://doi.org/10.1021/nn500277y.

(20) Pan, S.; Kong, W.; Liu, J.; Ge, X.; Zereshki, P.; Hao, S.; He, D.; Wang, Y.; Zhao, H. Understanding Spatiotemporal Photocarrier Dynamics in Monolayer and Bulk MoTe2 for Optimized Optoelectronic Devices. *ACS Appl. Nano Mater.* **2019**, *2* (1), 459–464. https://doi.org/10.1021/acsanm.8b02008.

(21) Palummo, M.; Bernardi, M.; Grossman, J. C. Exciton Radiative Lifetimes in Two-Dimensional Transition Metal Dichalcogenides. *Nano Lett.* **2015**, *15* (5), 2794–2800. https://doi.org/10.1021/nl503799t.

(22) Villegas, C. E. P.; Rocha, A. R. Elucidating the Optical Properties of Novel Heterolayered Materials Based on MoTe2–InN for Photovoltaic Applications. *J. Phys. Chem. C* **2015**, *119* (21), 11886–11895. https://doi.org/10.1021/jp5122596.

(23) He, J.; He, D.; Wang, Y.; Cui, Q.; Ceballos, F.; Zhao, H. Spatiotemporal Dynamics of Excitons in Monolayer and Bulk WS2. *Nanoscale* **2015**, *7* (21), 9526–9531. https://doi.org/10.1039/C5NR00188A.

(24) Huo, N.; Yang, Y.; Wu, Y.-N.; Zhang, X.-G.; Pantelides, S. T.; Konstantatos, G. High Carrier Mobility in Monolayer CVD-Grown MoS2 through Phonon Suppression. *Nanoscale* **2018**, *10* (31), 15071–15077. https://doi.org/10.1039/C8NR04416C.

(25) Rawat, A.; Jena, N.; Dimple; Sarkar, A. D. A Comprehensive Study on Carrier Mobility and Artificial Photosynthetic Properties in Group VI B Transition Metal Dichalcogenide Monolayers. *J. Mater. Chem. A* **2018**, *6* (18), 8693–8704. https://doi.org/10.1039/C8TA01943F.